%% file: sp26_main.tex
\documentclass[conference,compsoc]{IEEEtran}
\ifCLASSOPTIONcompsoc
  \usepackage[nocompress]{cite}
\else
  \usepackage{cite}
\fi

\ifCLASSINFOpdf
\else
\fi

\usepackage{tikz}
\usepackage{amsmath}
\usepackage{amsfonts}

\usepackage{algorithm}
\usepackage{algpseudocode}
\usepackage{xcolor}
\usepackage[skip=1pt]{subcaption}
\usepackage{longtable}
\usepackage{indentfirst}
\usepackage{dsfont}
\usepackage{bbm}
\usepackage{hyperref}
\usepackage{url} 
\usepackage{graphicx}
\usepackage{multirow}
\usepackage{pifont}
\usepackage{amsthm}
\usepackage{balance}
\usepackage{booktabs}
\usepackage{makecell}
\usepackage{colortbl}

\usepackage{tikz}
\usetikzlibrary{arrows.meta, positioning}
\usepackage{enumitem}
\setlist[itemize]{left=.1cm}
\setlist[enumerate]{left=.1cm}

\newtheorem{theorem}{Theorem}

\newtheorem{lemma}{Lemma}

\newtheorem{assumption}{Assumption}
\newtheorem{definition}{Definition}

\newcommand{\fmgftoc}{-0.1cm}
\newcommand{\fmgctom}{0cm}

\newcommand{\smgbs}{-0.3cm}
\newcommand{\MargBtParag}{0.05cm}

\newcommand{\equsize}{\small}
\newcommand{\equienvs}{\begin{normalsize}}
\newcommand{\equienve}{\end{normalsize}}

\newcommand{\BestfChannel}{$(1,\mathcal{M}_f, \mathbf{H}_f, \mathbf{D}^{opt}_f, 1)$}

\newcommand{\notbf}[1]{\noindent\textbf{#1}}
\newcommand{\E}[2]{\underset{#1}{\mathbb{E}}\left[#2\right]}
\newcommand{\prob}{\operatorname{Pr}}

\newcommand{\pr}[2]{{\ifx&#1& \prob \else \underset{#1}{\prob} \fi \left[#2\right]}}
\newcommand{\prc}[3]{{\ifx&#1& \prob \else \underset{#1}{\prob } \fi \left( #2 \middle| #3 \right)}}

\newcommand{\sumb}[2]{{\ifx&#1& \prob \else \underset{#1}{\sum} \fi #2}}

\newcommand{\RevisionChangeStart}{\begin{color}{black}}
\newcommand{\RevisionChangeEnd}{\end{color}}

\newcommand{\sheperdrevisionStart}{\begin{color}{black}}
\newcommand{\sheperdrevisionEnd}{\end{color}}

\begin{document}
\title{Tight Privacy Audit in One Run}

\author{
{\rm Zihang Xiang}\\
UCLA
\and
{\rm Tianhao Wang}\\
University of Virginia
\and
{\rm Hanshen Xiao}\\
Purdue University
\and
{\rm Yuan Tian}\\
UCLA
\and
{\rm Di Wang}\\
KAUST
% copy the following lines to add more authors
% \and
% {\rm Name}\\
%Name Institution
} % end author

\maketitle

\begin{abstract}
In this paper, we study the problem of privacy audit in one run and show that our method achieves tight audit results for various differentially private protocols. This includes obtaining tight results for auditing $(\varepsilon,\delta)$-DP algorithms where all previous work fails to achieve in any parameter setups. We first formulate a framework for privacy audit \textit{in one run} with refinement compared with previous work. Then, based on modeling privacy by the $f$-DP formulation, we study the implications of our framework to obtain a theoretically justified lower bound for privacy audit. In the experiment, we compare with previous work and show that our audit method outperforms the rest in auditing various differentially private algorithms. We also provide experiments that give contrasting conclusions to previous work on the parameter settings for privacy audits in one run. Our implementation is at \url{https://github.com/zihangxiang/tight-privacy-audit-in-one-run.git}

\end{abstract}

\begin{IEEEkeywords}
Privacy Audit, Statistical Estimation, Differential Privacy
\end{IEEEkeywords}

\input{sp26_content}

% %-------------------------------------------------------------------------------
% \section*{Acknowledgments}
% %-------------------------------------------------------------------------------
% %-------------------------------------------------------------------------------
% \section*{Availability}
% %-------------------------------------------------------------------------------
%-------------------------------------------------------------------------------
\bibliographystyle{plain}
\bibliography{ref}

\input{sp26_appendix}

% \balance

%%%%%%%%%%%%%%%%%%%%%%%%%%%%%%%%%%%%%%%%%%%%%%%%%%%%%%%%%%%%%%%%%%%%%%%%%%%%%%%%
\end{document}

%% file: sp26_content.tex
\section{Introduction}\label{sec:introduction}
\vspace{-0.2cm}
Differential privacy (DP) \cite{dwork2014algorithmic,dwork2009complexity,abadi2016deep} has emerged as a rigorous and principled framework for mitigating privacy risks in data analysis. By ensuring that the output of an algorithm remains statistically indistinguishable when any single data point is added or removed, DP provides robust, worst-case privacy guarantees. This is typically achieved through bounding the influence of any individual datapoint and adding calibrated randomness.

Unfortunately, because of the complexity of the underlying privacy analysis and the potential implementation bugs \cite{tramer2022debugging}, errors have been found in the use of classic DP building blocks like the Sparse Vector Technique, where flawed analysis leads to an algorithm that is not as private as claimed \cite{chen2015privacy}. 
And even if the algorithm's privacy analysis is indeed sound, unexpected errors can still arise during the deployment phase, such as in the presence of floating-point vulnerability \cite{mironov2012significance} or poor management of the seed to the pseudorandom generator \cite{PRNG_key_reuse}. These challenges suggest that purely analytical verification of DP is not sufficient in practice. 

To address this gap, empirical privacy auditing \cite{nasr2021adversary,nasr2023tight,steinke2023privacy,xiang2025privacy} has emerged as an effective approach for verifying the privacy of some algorithms $\mathcal{M}$ claimed to be differentially private. Privacy audit technique forms a binary hypothesis testing problem, and a privacy lower bound is concluded based on the result of hypothesis testing. While conceptually simple and powerful, one notable limitation of privacy audit is that it requires a significant number of runs of the DP algorithm (often at least thousands of runs) to reach non-trivial confidence. In a conventional audit approach, one hypothesis testing corresponds to one run of $\mathcal{M}$. This is problematic when $\mathcal{M}$ becomes bulky, e.g., training of a large language model (LLM)~\cite{cost_llm}.

To mitigate this issue, one heuristic has been proposed: performing multiple hypothesis testings on many data examples simultaneously while only running $\mathcal{M}$ once \cite{malek2021antipodes,zanella2023bayesian}. However, the dependencies between each hypothesis testing make it difficult to conclude a result with non-trivial confidence, as typical statistical estimation methods require samples to be independent. For example, the Clopper-Pearson method \cite{clopper1934use} used by Nasr et al. \cite{nasr2021adversary} to estimate the false positive rate (FPR) and false negative rate (FNR) of the hypothesis testing requires independent trials, but it is no longer applicable when each hypothesis testing is not independent of each other.

To mitigate this problem, a recent advancement \cite{steinke2023privacy} in privacy audit shows that it is possible to perform the hypothesis testing multiple times under a single run of $\mathcal{M}$, and a privacy lower bound can be produced with non-trivial confidence. 
%
% \vspace{\MargBtParag}
% \notbf{Motivation.} 
Several evaluations have been performed in \cite{steinke2023privacy}, but the tight result is only seen for $\mathcal{M}$ satisfying pure differential privacy ($(\varepsilon,0)$-DP). For the targeted algorithm $\mathcal{M}$ satisfying the more general $(\varepsilon,\delta > 0)$-DP, no tight result is achieved under any parameter setups. It is not clear what factors lead to such a gap.
This is also true for a later work \cite{mahloujifar2024auditing} that improves over \cite{steinke2023privacy}.

% \vspace{\MargBtParag}
\notbf{Contribution.} We show that it is possible to achieve tight privacy audit for the targeted algorithm $\mathcal{M}$ satisfying general privacy guarantees in one run. Our contributions are as follows:

\textit{\underline{A framework for privacy audit in one run}.} We formulate a general framework for privacy audit in one run (and it is also trivially applicable for privacy audit by multiple runs). In this framework, privacy audit is modeled as guessing secret bits: we have $n$ independent bits hidden in the execution of the targeted algorithm $\mathcal{M}$, and the goal is to guess these bits as accurately as possible purely based on the output of $\mathcal{M}$. 

Our framework has a specifically designed module that accounts for the heuristic of only releasing a subset of guesses proposed in \cite{steinke2023privacy}: we can associate scores with each guess and only release those guesses with high scores (i.e., those guesses are believed to be more accurate). Note that either $\mathcal{M}$ itself or the above heuristic will make guesses to be \textit{dependent} on each other, which is the main technical difficulty haunting privacy audit in one run.

We give theoretical analysis that bypasses such an issue. We first model the privacy algorithm by hypothesis-testing-based differential privacy ($f$-DP). This perspective allows us to capture the privacy of $\mathcal{M}$ more informatively. To reason the fundamental implications of our framework to provide privacy audit principles, we abstract $\mathcal{M}$'s privacy from its unimportant details to form a \text{base distribution pair}. This greatly simplifies further reasoning, and such a technical result generalizes to $\mathcal{M}$ satisfying any $f$-DP.

We provide further investigation into the heuristic of only releasing a subset of guesses, as we believe such behavior deserves deeper understanding and its implications are ignored by previous work \cite{steinke2023privacy,mahloujifar2024auditing,xiang2025privacy}. We find that it should be framed as an \textit{order statistics} problem. Combining this new finding and our modeling built on the \textit{base distribution pair}, we are able to give a tight privacy lower bound for privacy audit in one run.

\textit{\underline{Experiments}.} We provide results for auditing various common differential private algorithms in the experiments. We show that, given the same outcome of guessing secret input bits, our method outperforms the rest in all cases and reaches tight audit results. We also highlight the inherent limitation of the previous method by Steinke et al. \cite{steinke2023privacy} for auditing algorithms satisfying $(\varepsilon,\delta)$-DP, where their method can fail to give any meaningful results under certain parameter setups. 

Another notable audit experiment is provided for auditing the sub-sampled Gaussian mechanism \cite{abadi2016deep,rdp_subgaussian_mir}. This new layer of randomness due to sub-sampling brings additional difficulties to the privacy audit. However, we show in the experiment that we can still achieve tight audit results for the sub-sampled Gaussian mechanism in one run. In contrast, previous methods can only give very weak results.

We also give discussions regarding the heuristic of only releasing a subset of guesses. By ablation-studying the audit performance under different numbers of released guesses, we show contrasting results between our method and previous work \cite{steinke2023privacy,mahloujifar2024auditing}. Unlike previous work, the results show evidence that our method is not limited to the need for a trade-off in determining the number of guesses to be released.

For privacy audit in one run, our work can be used as an add-on to give improved audit results, given a collection of guesses.

\section{Preliminaries}\label{sec:preliminaries}

\subsection{Differential Privacy (DP)} 
% \vspace{\pmgttom}
\vspace{-0.2cm}
\begin{definition}[Differential Privacy \cite{DworkMNS06}]\label{def:dp}
Let $\mathcal{M}: \mathcal{X}^* \to \mathcal{Y}$ be a randomized algorithm, where $\mathcal{X}^* = \bigcup_{n \ge 0} \mathcal{X}^n$.
    We say $\mathcal{M}$ is $(\varepsilon,\delta)$-differentially private ($(\varepsilon,\delta)$-DP) if, for all $X,X' \in \mathcal{X}^*$ differing only by one element, we have $\forall S \subset \mathcal{Y}$ 
    $$\prob(\mathcal{M}(X) \in S) \le e^\varepsilon \cdot \prob(\mathcal{M}(X') \in S) + \delta.$$
\end{definition}
In the context of privacy audit, we care about how hard it is to distinguish one distribution from the other based on a single draw. The above definition of DP requires that two distributions $\mathcal{M}(X)$ and $\mathcal{M}(X')$ are close to each other, which makes them hard to distinguish from each other.

\vspace{\MargBtParag}
\notbf{Hypothesis-testing based differential privacy: $f$-DP.} 
$f$-DP formulation mitigates the limitation of $(\varepsilon,\delta)$-DP by characterizing the privacy of a private algorithm $\mathcal{M}$ with a function $f$ rather than a single pair of parameters $(\varepsilon,\delta)$. The \textit{trade-off function} $f$ is defined as the best Type II error $\beta$ (a.k.a., false negative rate) one can achieve for fixed Type I error $\alpha$ (a.k.a., false positive rate) in a hypothesis testing problem \cite{dong2019gaussian}. The setting is as follows. For distributions $P$ and $Q$ on domain $\mathcal{Y}$, define

\begin{equation}\label{equ:basic_hypo}
% \equsize
\begin{aligned}
H_0:P,\text{\quad}H_1:Q. 
\end{aligned}
\end{equation}
% \equienve
For any decision rule (also called as a \textit{test}) $\mathcal{R}:\mathcal{Y}\rightarrow \{0,1\}$, one must trade off $\alpha$ and $\beta$ which are defined as follows:
\begin{equation}\label{equ:basic_hypo_error}
    \alpha=\pr{}{\mathcal{R}(y)=1|H_0},\text{\quad}\beta=\pr{}{\mathcal{R}(y)=0|H_1}.
\end{equation}
We give the definition of the trade-off function in the following.
\begin{definition}[Trade-off Function]\label{def:trade_off_function}
    A trade-off function $T_{P,Q}:[0,1]\rightarrow[0,1]$ for some distribution $P,Q$ is defined as
\begin{equation}\label{equ:trade_off_function}
    T_{P,Q}(\alpha)=\inf_{\mathcal{R}}\{\beta_{\mathcal{R}}:\alpha_\mathcal{R}\leq \alpha\}
\end{equation}
\end{definition}
It is worth noting that the trade-off function $T_{P,Q}$ is convex, continuous, and non-increasing \cite{dong2019gaussian}. The best $\beta$ for fixed $\alpha$ can always be obtained via the well-known likelihood ratio test, also known as the Neyman-Pearson Lemma \cite{neyman1933ix}. The likelihood ratio is related to the privacy loss distribution (PLD) regarding $P$ and $Q$ defined below.
\sheperdrevisionStart
\begin{definition}[Privacy Loss Distribution (PLD) \cite{zhu2022optimal}]
    \label{def:pld} 
    For two continuous distributions $P$ and $Q$, their privacy loss at $o$ is defined as
    \begin{equation}\label{equ:plrv}
        \mathcal{L}_{P / Q}(o):=\ln \left(\frac{P(o)}{Q(o)}\right)
    \end{equation}
    where $P(o)$ and $Q(o)$ are the probability density functions of $P$ and $Q$ at point $x$, respectively.
    The Privacy Loss Distribution (PLD) of $P$ and $Q$, denoted by $P L D_{P / Q}$, is a distribution on $\mathbb{R} \cup\{\infty\}$ where $y \sim P L D_{P / Q}$ is generated by sampling $o \sim P$ and then letting $y=\mathcal{L}_{P / Q}(o)$.
\end{definition}
\sheperdrevisionEnd

For a randomized algorithm $\mathcal{M}$ under adjacent datasets $X,X'$ the privacy loss distribution (PLD) is $PLD_{\mathcal{M}(X)/\mathcal{M}(X')}$.
For two discrete distributions, the privacy loss is defined similarly, i.e., just replace the probability density function with the probability mass function.
    
As we can see, the privacy loss at some point is the log-likelihood ratio. Using the trade-off function, we can quantify the privacy of a private algorithm $\mathcal{M}$ by the following definition.
\begin{definition}[$f$-DP \cite{dong2019gaussian}]
    Let $f:[0,1]\rightarrow [0,1]$ be a trade-off function. A mechanism $\mathcal{M}$ is f-DP if 
    % \equienvs
    % \large
    \begin{equation}\nonumber
        T_{\mathcal{M}(X),\mathcal{M}(X')}(y)\geq f(y), \forall y\in[0,1]
    \end{equation}
    holds for all adjacent dataset pairs $X, X'$.
\end{definition}
Suppose $f = T_{P,Q}$ for some pair of distribution $P, Q$,  $\mathcal{M}$ is $f$-DP means that distinguishing $\mathcal{M}(X)$ from $\mathcal{M}(X')$ is at least as hard as distinguishing $P$ from $Q$. $f$-DP is a generalization of $(\varepsilon,\delta)$-DP formulation \cite{dong2019gaussian,wasserman2010statistical}. Specifically, the following two statements are equivalent \cite{zhu2022optimal}: 1) $\mathcal{M}$ is $(\varepsilon,\delta)$-DP; 2) $\mathcal{M}$ is $f_{\varepsilon,\delta}$-DP where the trade-off function $f_{\varepsilon,\delta}$ is 
$$ f_{\varepsilon,\delta}(y)=\max{(0,1-\delta-\mathrm{e}^\varepsilon y,\mathrm{e}^{-\varepsilon}(1-\delta-y))}$$

The privacy of the Gaussian mechanism is captured by the indistinguishability between two Gaussians $\mathcal{N}(0,1)$ and $\mathcal{N}(\mu,1)$ \cite{dong2019gaussian}. The following family of trade-off functions quantifies the privacy of the Gaussian mechanism \cite{dong2019gaussian}, i.e., each point on the trade-off function (the best $\alpha,\beta$ pairs) is achievable.
\begin{definition}[$\mu$-Gaussian DP ($\mu$-GDP) \cite{dong2019gaussian}]\label{def:gdp}
    The trade-off function of distinguishing $\mathcal{N}(0,1)$ from $\mathcal{N}(\mu,1)$ is
    {$$ G_\mu(x)=T_{\mathcal{N}(0,1),\mathcal{N}(\mu,1)}(x)=\Phi(\Phi^{-1}(1-x)-\mu),$$}where $\Phi$ is the c.d.f. of the standard normal distribution. A private mechanism $\mathcal{M}$ satisfies $\mu$-GDP if it is $G_\mu$-DP
\end{definition}

Similarly, for the Laplace mechanism that adds Laplace noise instead of Gaussian noise, its privacy is captured by the indistinguishability between two Laplace distributions $\operatorname{Lap}(0,1)$ and $\operatorname{Lap}(\mu,1)$ where $\operatorname{Lap}(\mu,c)$ represents the Laplace distribution with mean $\mu$ and scale parameter $c$. Specifically,

\begin{lemma}[Laplace DP \cite{dong2019gaussian}]\label{lem:laplace_dp}
The trade-off function for distinguishing $\operatorname{Lap}(0, 1)$ from $\operatorname{Lap}(\mu, 1)$ is

\begin{equation}\label{equ:laplace_dp}
\begin{aligned}
    T_{\operatorname{Lap}(0,1), \operatorname{Lap}(\mu, 1)}(x)= \begin{cases}1-\mathrm{e}^\mu x, & x<\mathrm{e}^{-\mu} / 2 \\ \mathrm{e}^{-\mu} / 4 x, & \mathrm{e}^{-\mu} / 2 \leq x \leq 1 / 2 \\ \mathrm{e}^{-\mu}(1-x), & x>1 / 2\end{cases}
\end{aligned}
\end{equation}
\end{lemma}

\subsection{Privacy Audit} 
Empirical privacy audit gives a privacy lower bound associated with a confidence specification. For example, at some fixed $\delta$ and for some algorithm claiming to satisfy $(\varepsilon_u,\delta)$-DP, suppose its true privacy parameter is actually $\varepsilon_t$, a privacy audit gives a lower bound $\varepsilon_l\leq\varepsilon_t$. If $\varepsilon_l>\varepsilon_u$, it suggests the claimed privacy guarantee may not hold.

\begin{algorithm}[!ht]
\caption{
Distinguishing Game
}\label{alg:typical_audit}
\begin{algorithmic}[1]
% \equsize
\renewcommand{\algorithmicrequire}{\textbf{Input:}}
\renewcommand{\algorithmicensure}{\textbf{Output:}}

\Require {$\mathcal{M}$, targeted Algorithm;  $X, X'$, adjacent datesets; decoder $\mathbf{D}:\mathcal{Y}\rightarrow \{0,1\}$}
\State $b_{t} \gets \mathbf{Bernoulli}(1/2)$ 
\State $y\gets \mathcal{M}(X)$ if $b_{t}=0$, $y\gets \mathcal{M}(X')$ otherwise
\State $b_{g} \gets \mathbf{D}(y)$
\Ensure $(b_{t}, b_{g})$

\end{algorithmic}
\end{algorithm}

Typical workflow for implementing a privacy audit is summarised in Algorithm~\ref{alg:typical_audit}. It starts by sampling a random bit $b_t\gets \mathbf{Bernoulli}(1/2)$, then the sampled bit indicates a selection between two adjacent datasets $X$ and $X'$; then the targeted algorithm $\mathcal{M}$ is run on the selected dataset and outputs $y$; a final guessed bit $b_g$ is produced by a decoder $b_g=\mathbf{D}(y)$. 

Based on all the truths V.S. the guesses of this distinguishing game (simulated many times), we can make statistical estimations. For instance, Nasr et al. \cite{nasr2021adversary} compute FPR $\alpha$ and FNR $\beta$ with confidence by the Clopper-Pearson method \cite{clopper1934use}. Let $\alpha_r$ and $\beta_r$ be the rightside value of the confidence interval for estimating $\alpha$ and $\beta$, respectively, the privacy lower bound is computed by
\begin{equation}\label{equ:lower_bound}
% \equsize
\begin{aligned}
    \varepsilon_l=\max
    \{\log \frac{1-\delta-\alpha_r}{\beta_r}, \log \frac{1-\delta-\beta_r}{\alpha_r}, 0\}
\end{aligned}
\end{equation}

\section{Related Work}\label{sec:related_work}
The goal of a privacy audit is to provide refutation for the privacy guarantee of some targeted algorithm claimed to be differentially private. Some previous related work focuses on finding algorithm implementation bugs using program analysis methods \cite{farina2020coupled,wang2020checkdp,barthe2020deciding}. However, the limitation is that they mainly target algorithms satisfying pure differential privacy (pure DP) \cite{dwork2014algorithmic}. Other notable work \cite{bichsel2021dp,bichsel2018dp} converts the audit problem to a search problem by finding the output of the algorithm that serves to be the evidence of a privacy violation; however, this method is not applicable when running the targeted algorithm is expensive.

In the context of machine learning, the algorithm's privacy is probed by usually performing membership inference attacks (MIAs) \cite{shokri2017membership}; this technique is designed to assert the presence or absence (membership) of some data examples by inspecting the algorithm's output. Once the statistical validity of FPR and FNR is ensured, it gives a lower bound for the true privacy parameter $\varepsilon_t$. For example, Nasr et al. \cite{nasr2021adversary,nasr2023tight} model the FPR and FNR as two unknown parameters for two Bernoulli random variables, and they are estimated with confidence over \textit{independent} trials. Several improvements have been made to different stages of the privacy audit.

Jagielski et al. \cite{jagielski2020auditing} perform MIA on data examples (a.k.a. canaries) that are specifically poisoned. Cebere et al. \cite{cebere2024tighter} design contrived canaries to the hidden state model of DP-SGD and show that hiding the intermediate state of DP-SGD does not enhance its privacy guarantee. Maddock et al. \cite{maddock2022canife} construct canaries to understand the privacy lower bound in the federated learning settings. 
% To gain better MIA results, it is often necessary to design canaries with highly detectable features.

Other related works focus on how to better interpret the results of MIA to reach better statistical estimation. This involves better statistical analysis, e.g., Log-Katz confidence intervals \cite{lu2022general} and Bayesian techniques \cite{zanella2023bayesian} are proposed to give a stronger privacy lower bound. Nasr et al. \cite{nasr2023tight} consider the specific privacy property of the private algorithms to reach a tight audit result. Xiang et al. \cite{xiang2025privacy} also propose a better transformation from MIA results to the privacy lower bound based on mutual information.

One common property of the approaches to extract the privacy lower bound from the MIA result is that they require the FPR and FNR to be computed based on independent samples. To achieve this, a default choice is to run the targeted algorithm $\mathcal{M}$ multiple times independently, and each run serves as a Bernoulli trial. When $\mathcal{M}$ is expensive to run, one heuristic solution is to perform multiple MIA simultaneously in only one run of $\mathcal{M}$ \cite{malek2021antipodes,zanella2023bayesian}; however, the possible dependencies between the samples bring difficulties to the statistical analysis of the MIA result.

To address this issue, Steinke et al. \cite{steinke2023privacy} propose a general audit framework that allows performing multiple MIA simultaneously without re-running the targeted algorithm $\mathcal{M}$. The final concluded privacy lower bound with its justified confidence is related to how accurate the guesses are in MIA. Another work \cite{mahloujifar2024auditing} improves over \cite{steinke2023privacy} by considering the specific properties of the targeted algorithm $\mathcal{M}$. 

One notable fact is that both \cite{steinke2023privacy} and \cite{mahloujifar2024auditing} cannot give tight results for Gaussian mechanism even in the idealized setting. A later work by Xiang et al. \cite{xiang2025privacy} discusses such an issue and provides several tighter results. However, as shown by \cite{xiang2025privacy}, although it is not necessary to run the algorithm $\mathcal{M}$ multiple times, their tight results are still \textit{only valid when the results of each MIA are independent} from each other; the curse of dependency still remains outstanding for \cite{xiang2025privacy}.

One common trait shared by \cite{steinke2023privacy,mahloujifar2024auditing,xiang2025privacy} and our work is that the improvement in MIAs that provide better guesses can help to reach better audit results, but this is orthogonal to all of these works.

\section{Motivation \& Method Overview}\label{sec:Motivation}
\vspace{-0.2cm}
For interpreting MIA results over independent samples, previous work already closes the gap between the privacy lower bound and the upper bound \cite{nasr2023tight,nasr2021adversary}, showing that the privacy analysis is tight and cannot be improved further. In contrast,  for privacy audit in one run where possible dependencies exist, results given by \cite{steinke2023privacy} and \cite{mahloujifar2024auditing} show that there is still a significant gap between the lower bound and the upper bound. For instance, for an idealized setting in auditing the basic Gaussian mechanism with the upper bound known to be tight, neither \cite{steinke2023privacy} nor \cite{mahloujifar2024auditing} could close the gap.

Because the upper bound is tight, it must be that the gap is due to the suboptimality of the lower bound. However, it is unclear what causes the suboptimality. Is it because the audit framework is ineffective? Or is it because the dependencies between the results of MIA make the suboptimality fundamentally unavoidable?

This motivates us to study the privacy audit problem in one run, where any possible dependencies may exist between the results of each MIA, and our method gives positive answers that a tight audit is indeed achievable. 

% \vspace{\MargBtParag}
\notbf{Method overview.} The final privacy lower bound is concluded based on the contrapositive of differential privacy: if the targeted algorithm $\mathcal{M}$ is indeed differentially private, then it is impossible to guess the bits very accurately. Therefore, if we have some accurate guesses, it suggests that $\mathcal{M}$ does not satisfy some privacy guarantee. However, in the context of privacy audit in one run, converting guesses to a privacy lower bound requires significant technical manipulations.

We first give a framework for modeling the general behaviors of privacy audit in one run. Specifically, $n$ independent bits are sampled, and each bit indicates whether a data example is in the input dataset $X_{in}$ or not. The targeted algorithm $\mathcal{M}$ is run on $X_{in}$, and the output of $\mathcal{M}$ is fed to a decoder $\mathbf{D}$ that guesses the bits based on the output. The decoder $\mathbf{D}$ also associates a score with each guess. Finally, only $r$ guesses with the highest scores are released. The final privacy lower bound is concluded based on the released guesses and their corresponding true bits. There are several notable features in our analysis of our framework:

\begin{enumerate}
    % \vspace{-0.3cm}
    \item We model the target algorithm $\mathcal{M}$'s privacy by $f$-DP. Although $f$-DP auditing has been studied previously \cite{mahloujifar2024auditing,xiang2025privacy}, we are the first to combat the gap between the privacy lower bound and the upper bound in one run, while previous work can not: \cite{mahloujifar2024auditing} still cannot close the gap for auditing Gaussian mechanism in the idealized setting, and \cite{xiang2025privacy}'s result is not applicable in general audit-in-one-run cases, as guesses are dependent.

    % \vspace{-0.2cm}
    \item We reduce the privacy of $\mathcal{M}$ to be a \textit{base distribution pair} that captures the fundamental implication of the privacy of $\mathcal{M}$, and we show that the base distribution pair is sufficient for our purpose. The reduction in quantifying privacy hides unimportant details and allows for more manageable theoretical analysis.
    Our approach also generalizes to the case where $\mathcal{M}$ satisfies any $f$-DP by construction.

    % \vspace{-0.2cm}
    \item We show that the heuristic of only releasing a subset of guesses based on their scores should be modeled by \textit{order statistics}. We also argue that this modeling captures the implication behind the actions of releasing a subset of guesses. This is a new finding that has not been discussed in previous work \cite{steinke2023privacy,mahloujifar2024auditing,xiang2025privacy}. How \textit{order statistics} fits into our framework is formally quantified.
\end{enumerate}
With the above features combined, we are able to convert guesses to a tight privacy lower bound for privacy audit in one run. 

% \vspace{\smgbs}
\section{Methodology}\label{sec:methodology}
\subsection{Audit Framework}\label{sec:audit_principle}
\begin{figure}[!ht] 
    \centering
    % \subfloat[]
    {
    \includegraphics[width=0.95\linewidth]{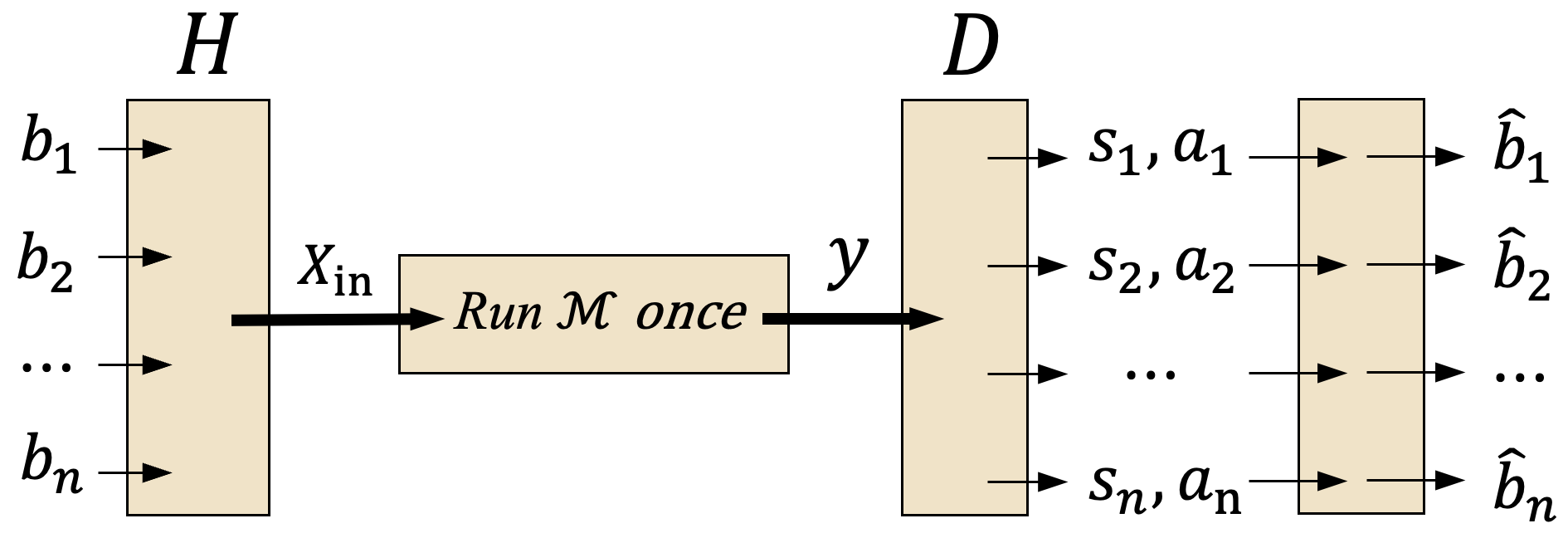}
    }
    % \vspace{\fmgftoc}
    \caption{
    Our audit framework. Dataset $X_{in}$ is formed based on the input bits $\{b_i:i\in[n]\}$. The targeted privacy algorithm $\mathcal{M}$ is only run for a single time. The output of $\mathcal{M}$ is fed to a decoder $\mathbf{D}$ where a score and a guessed bit pair are formed for each input bit. Finally, only $r$ guessed bits are released based on their associated scores.
    }
    \label{fig:audit_frame} 
    % \vspace{\fmgctom}
\end{figure}

\vspace{-0.2cm}
We first introduce our formulation for the privacy audit problem. Our formulation is a combination and refinement of the audit framework provided in \cite{steinke2023privacy} and \cite{xiang2025privacy}. We also borrow some useful naming and notation conventions from \cite{xiang2025privacy}. At a high level, in our framework, we have some independent secret bits to be guessed based on the output of the targeted private algorithm that runs only once. Our framework is described in the following definition and is pictorially presented in Figure \ref{fig:audit_frame}.

\begin{definition}[General Audit Framework]\label{def:general_audit_framework}
    The general privacy audit framework is a tuple $(n, \mathcal{M}, \mathbf{H}, \mathbf{D}, r)$ consisting of the following components:

    \begin{enumerate}
        % \vspace{-0.2cm}
        \item $n$: number of bits to be guessed where each bit $b_i, \forall i\in[n]$ is independently sampled from $\mathbf{Bernoulli}(1/2)$.
        % \vspace{-0.2cm}
        \item $\mathcal{M}:\mathcal{X}^* \to \mathcal{Y}$: the targeted 
        differentially private algorithm.
        % \vspace{-0.2cm}
        \item $\mathbf{H}$: forming input dataset $X_{in}$ to be fed to the targeted private algorithm $\mathcal{M}$ based on input bits. 
        % \vspace{-0.2cm}
        \item $\mathbf{D}:\mathcal{Y}\rightarrow \mathbb{R}^{\geq 0}\times\{0,1\}$: decoder $\mathbf{D}$, based on the output $y$ of $\mathcal{M}$, computes guesses $a_i\in\{0,1\}$ and the corresponding scores $s_i\in\mathbb{R}^{\geq 0}$ for each input bit.
        % \vspace{-0.2cm}
        \item $r\in[n]$: The final filtering action releases $r$ guessed bits based on scores $\{s_i:i\in[n]\}$.  
        % Only releases $r$ guessed bits among $\{a_i:i\in[n]\}$ with the highest $r$ scores. 
        Specifically, suppose $\mathbb{S}_r$ is the set of scores of released bits and $\mathbb{S}_d$ is the set of scores of discarded bits, the filtering action ensures that:
        \begin{equation}\label{equ:filter_condition}
        % \large
            \min(\mathbb{S}_r) \geq \max(\mathbb{S}_d)
        \end{equation} 
        For released result, $\hat{b}_i = a_i$ if $i$-th bit is released, otherwise $\hat{b}_i = \bot$.
        % \item $r\in[n]$: the number of \underline{r}emaining valid guessed bits, i.e., $r = |\{i: \hat{b}_i\neq \bot\}|$.
    \end{enumerate}
\end{definition}

\notbf{Formulation highlight: dataset.} In addition to $X_{in}$, which contains data examples (also called \textit{canaries} in literature \cite{nasr2023tight,steinke2023privacy}) with membership to be guessed, $\mathcal{M}$ may also contain a ``hard-coded'' dataset that is for normal training. For example, the ``hard-coded'' dataset can be the original training dataset for image classification; $X_{in}$ can be some images specifically designed such that their membership is believed to be more easily determined. $\mathcal{M}$ takes input the union of $X_{in}$ and the ``hard-coded'' dataset. The ``hard-coded'' dataset is of less interest in the context of a privacy audit.

\notbf{Formulation highlight: filtering.} Inspired by the abstaining guessing action in \cite{steinke2023privacy}. We introduce a new formulation of the final \textit{filtering} action to allow the additional flexibility of only releasing a subset (not all of the guessed bits) based on their scores (we propose to compute it by the likelihood ratio). The high-level intuition is that if it is believed some bit guesses are ``bad'' (with low scores), it is allowed that they may not be released, as they could lead to a more trivial privacy lower bound. For example, random or low-confidence guessing is of no help and should always be discarded. Our formulation allows quantitative reasoning as will be seen.

% \vspace{\MargBtParag}
\notbf{Notations.} The following sections depend on the notations shown in Table \ref{tab:notations}. For denoting the realization of some random variable, we use its corresponding lowercase. When there are no ambiguities, we abuse the uppercase to denote its distributions.

\begin{table}[!ht] 
\equsize
% \vspace{-0.1cm}
\centering
% \begin{subtable}[h]{\textwidth}
% \raggedleft
% \centering
\resizebox{.95\columnwidth}{!}{
% \left
\begin{tabular}{ll}
% \toprule
\toprule

\multicolumn{2}{c}{Random Vector}\\
\midrule
$B$ & Random bit vector input to $\mathbf{H}$\\
$A$ & Random bit vector guessed by decoder $\mathbf{D}$\\
$\hat{B}$ & Random vector after the filtering\\
$E$ & Bit error random vector (Equation \eqref{equ:bit_error_rv})\\
\midrule
\multicolumn{2}{c}{Indexing conventions}\\
\midrule
$R_{i}$ & $i$-th coordinate of vector R\\
$R_{-i}$ & $[R_1, \cdots, R_{i-1}, R_{i+1},\cdots R_{n}]$\\
$R_{<i}$ & $[R_1, \cdots, R_{i-1}]$\\
$R_{>i}$ & $[R_i+1, \cdots, R_{n}]$\\
\midrule
\multicolumn{2}{c}{Others}\\
\midrule
$\mathcal{M}$ & Targeted algorithm to be audited\\
$y$ & Output of $\mathcal{M}$ in one run\\
$P,Q$ & Base distribution pair (Definition \ref{def:tight_distribution_pair})\\
$P(x),Q(x)$ & The density at point $x$\\
$T_{P,Q}$ & Trade-off function for distribution $P,Q$)\\
$R|B=b$ & Distribution of $R$ conditioned on $B=b$\\
$\preceq$ & Transmission dominance (Definition \ref{def:trans_dominance})\\
$\mathbf{D}$ & Decoder producing scores and guesses\\
$\mathbf{D}_f$ & Decoder in $f$-channel (Definition \ref{def:f_dp_channel})\\
$\mathbf{D}^{opt}_f$ & Optimal decoder of $\mathbf{D}_f$\\
$\mathcal{L}_{P/Q}$ & Privacy loss for $P,Q$ (Definition \ref{def:pld})\\
$S_{(k)}$ & $k$-th order statistic (Definition \ref{def:k-th_order_stats_def}) \\
$\mathbf{event}_u$ & Event defined in Equation \eqref{equ:event_u}\\
$\mathbf{H}$ & Defined in Definition \ref{def:general_audit_framework}\\
\bottomrule
% \caption{accuracy}
\end{tabular}

}

\caption{
\sheperdrevisionStart 
Notation used. 
\sheperdrevisionEnd 
}
\label{tab:notations}
% \vspace{-0.5cm}
\end{table}

\notbf{Central properties of the audit framework.} To deal with scenarios where $\mathcal{M}$ is DP with respect to add/removal adjacency, $b_i$ indicates whether to include or not include a certain data example. Naturally, for replacement adjacency, $b_i$ indicates a selection between two candidate canaries. In either case, $\mathcal{M}$'s output $y$ is DP with respect to the flip of any input bit. This is also the fundamental constraint due to DP underlying our framework:

Let $Y|{B_i=b_i, B_{-i}=b_{-i}}$ be $\mathcal{M}$'s output distribution conditioned on $B_i=b_i, B_{-i}=b_{-i}$ and  $Y|{B_i=1-b_i, B_{-i}=b_{-i}}$ for conditioning on ${B_i=1-b_i, B_{-i}=b_{-i}}$. Given that $\mathcal{M}$ is $f$-DP, 
\begin{large}
\begin{equation}\label{equ:dp_condtrain}
\begin{aligned}
    &T_{Y | {b_i, B_{-i}=b_{-i}}, \hspace{0.1cm}Y| {1-b_i, B_{-i}=b_{-i}}}\geq f\\
\end{aligned}
\end{equation}
\end{large}
$\forall i\in[n], b_{i}\in\{0,1\}, b_{<i}\in\{0,1\}^{n-1}$, by the definition of $f$-DP. 

\begin{definition}\label{def:all_error}
    Define a random variable $E_i$ as follows. 
    \begin{equation}\label{equ:bit_error_rv}
        E_i = B_i \oplus A_i
    \end{equation}
    where $\oplus$ is $\mathbf{XOR}$ operation. $E_i$ is a random variable that indicates whether the bit guessed by the decoder $\mathbf{D}$ is correct or not.
\end{definition}
It should be highlighted that $E_i$ is only known to us if the corresponding guessed bit is released, i.e., $\hat{b}_i\neq \bot$. Regarding the performance in making errors, we define the following relation between two audit frameworks.

\begin{definition}[Transmission Dominance]\label{def:trans_dominance}
    We say audit framework $(n, \mathcal{M}, \mathbf{H}, \mathbf{D}, r)$ is transmission-dominated by another audit framework $(n,\mathcal{M}', \mathbf{H}', \mathbf{D}', r)$ if the following holds:
    \begin{equation}\label{equ:dominance_relation}
    % \equsize
        \pr{B,\mathcal{M}}{\mathbf{event}_u} \leq \pr{B,\mathcal{M}'}{\mathbf{event}_u}, \forall u \in [n]
        \end{equation}
        where event $\mathbf{event}_u$ is defined as: 
        \begin{equation}\label{equ:event_u}
            \mathbf{SUM}(\{E_i: \hat{B}_i\neq \bot\})\leq u,
        \end{equation}
        i.e., for $r$ guessed bits, at most $u$ bits are guessed incorrectly. We express such a relation as: 
        \begin{equation}\label{equ:dominance_relation_succinct}
            (n, \mathcal{M}, \mathbf{H}, \mathbf{D}, r) \preceq (n,\mathcal{M}', \mathbf{H}', \mathbf{D}', r)
    \end{equation}

\end{definition}
Intuitively, Equation \eqref{equ:dominance_relation_succinct} states the fact that the audit framework $(n, \mathcal{M}, \mathbf{H}, \mathbf{D}, r) $ gives better results than $(n, \mathcal{M}', \mathbf{H}', \mathbf{D}', r) $ in terms of achievable bit errors. 

\subsection{Privacy Reduction}\label{sec:Reduction}
\begin{figure}[!t] 
    \centering

    \subfloat[$1$-GDP]
    {
    \includegraphics[width=.95\linewidth]{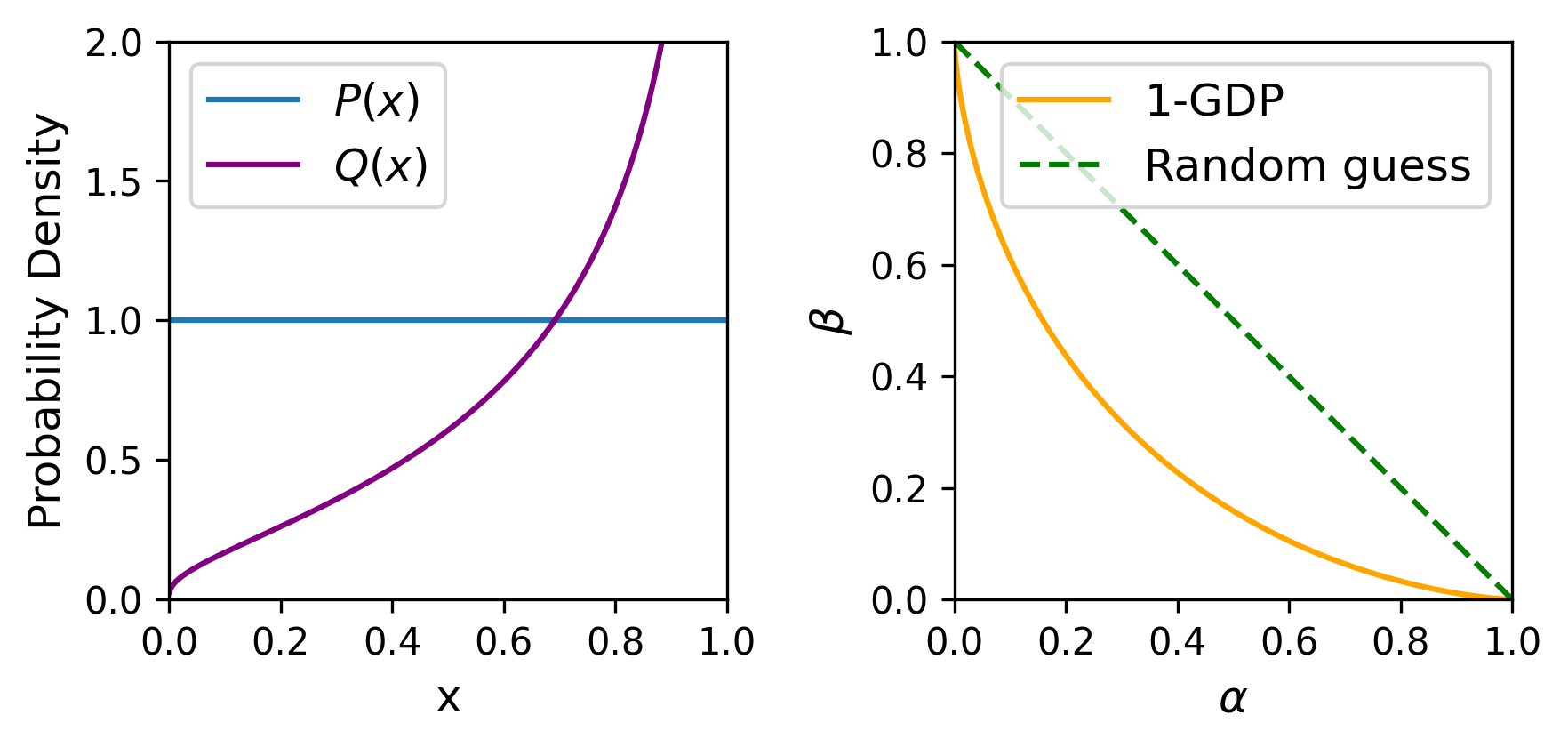}
    }

    \subfloat[$T_{\operatorname{Lap}(0,1),\operatorname{Lap}(\mu,1)}$]
    {
    \includegraphics[width=.95\linewidth]{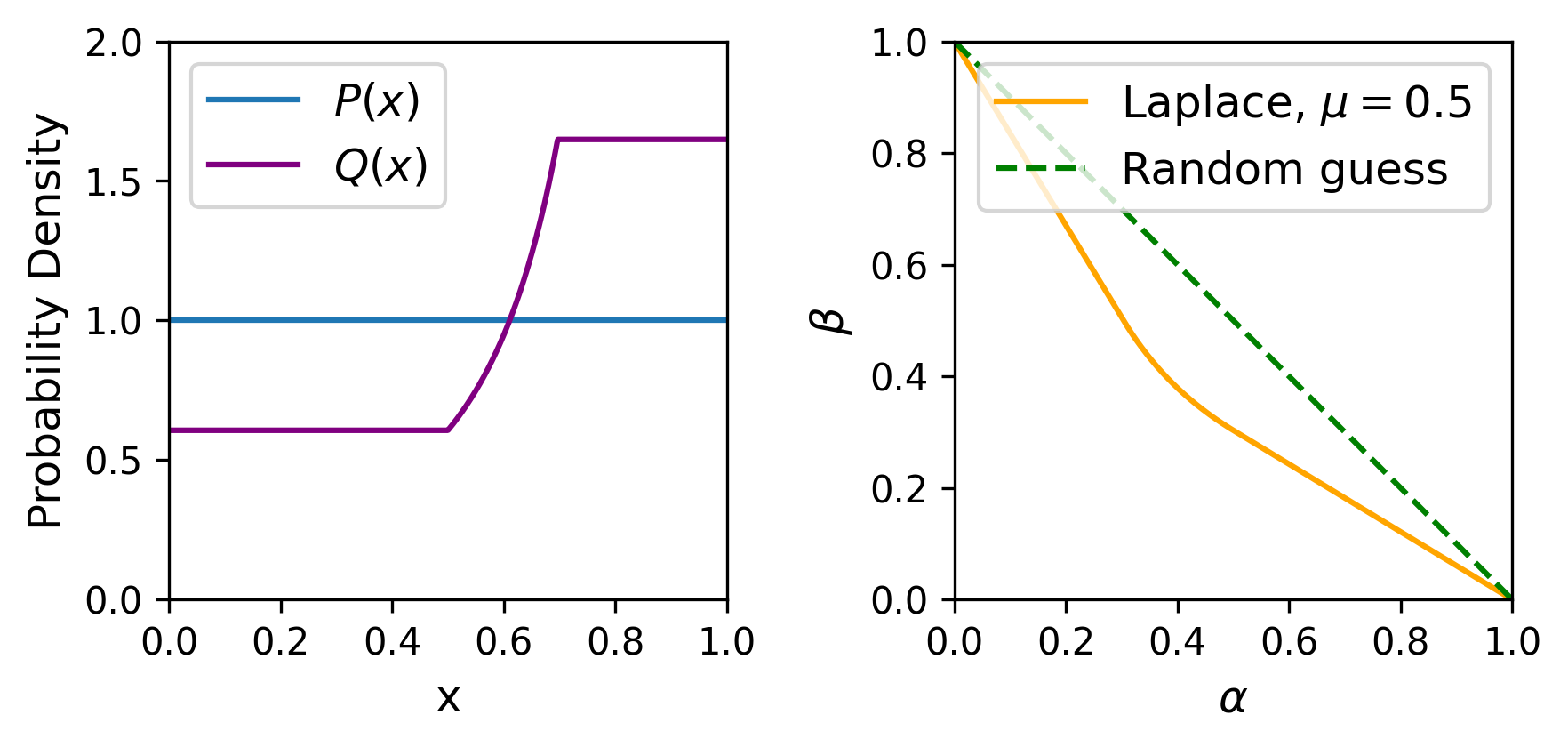}
    }
    
    \subfloat[$(0.5,0.1)$-DP]
    {
    \includegraphics[width=.95\linewidth]{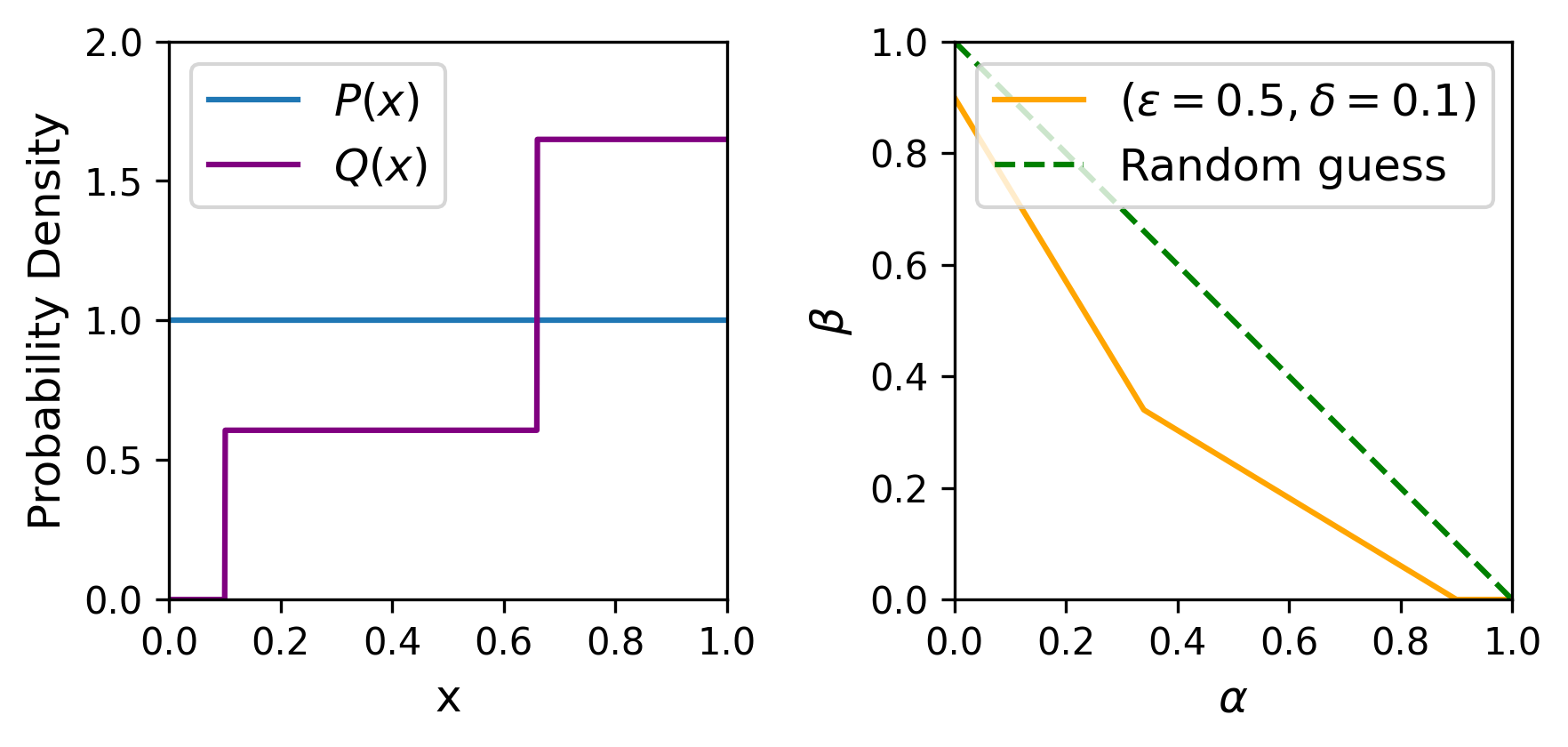}
    }
    % \subfloat[$\mu=3.2$]
    % {
    % \includegraphics[width=.33\linewidth]{figures_tables/n_vs_eps_lb_gdp3.2.pdf}
    % }
    % \vspace{\fmgftoc}
    \caption{
    The base distribution pairs $P,Q$ for different trade-off functions $f$.
    }
    \label{fig:dif_PQ} 
    % \vspace{\fmgctom}
\end{figure}

\vspace{-0.2cm}
Recall when we refer to $\mathcal{M}$ satisfying $f$-DP, we mean $T_{\mathcal{M}(X),\mathcal{M}(X')}\geq f$ for all adjacent datasets $X,X'$. To quantify the privacy of $\mathcal{M}$, we are in fact only interested in quantifying the indistinguishability between distributions $\mathcal{M}(X)$ and $\mathcal{M}(X')$. To serve this purpose and to make further reasoning easier, we introduce the following special pair of distributions $P,Q$ such that the trade-off function $T_{P,Q}=f$.
\begin{definition}[Base Distribution Pair]\label{def:tight_distribution_pair}
    For some algorithm $\mathcal{M}$ satisfying $f$-DP, a pair of distributions $P,Q$ is called a base distribution pair for $\mathcal{M}$ if the trade-off function $T_{P,Q}$ is exactly equal to the trade-off function $f$, i.e., $T_{P,Q}=f$. Specifically, $P$ and $Q$ can be constructed as follows:

    The domain is on the unit interval $[0,1]$. $P$ is the uniform distribution on $[0,1]$ and $Q$ has the following density: 
    $$Q(x) = -f'(1-x). $$ And $Q$ has point mass $1-f(0)$ at point $1$. The idea behind such construction is to let the c.d.f. of $Q$ be $f(1-x)$. For some point $x\in(0,1)$ such that $f$ is not differentiable, $Q(x)$ is arbitrarily set to be any subgradient of $f$ at $x$ (this is valid for our later use because the integral over such ``singular'' points is always zero).
\end{definition}
The above $P,Q$ construction is due to Dong et al. \cite{dong2019gaussian}. And we present several constructed examples in Figure \ref{fig:dif_PQ}. This reduction lets us focus on the most important part and reason on auditing the privacy of $\mathcal{M}$ more easily without loss of generality. Specifically, given
$$
T_{\mathcal{M}(X),\mathcal{M}(X')}\geq T_{P,Q}
$$
there exists a (randomized) post-processing \cite{dong2019gaussian} $\mathbf{Proc}:[0,1]\rightarrow \mathcal{Y}$ such that 
\begin{equation}\nonumber
\begin{aligned}
    &\mathbf{Proc}(P)=\mathcal{M}(X)\\
    &\mathbf{Proc}(Q)=\mathcal{M}(X')
\end{aligned}
\end{equation}
Post-processing only makes the distinguishing game harder. What cannot be achieved in distinguishing $P$ from $Q$, also cannot be achieved in distinguishing $\mathcal{M}(X)$ from $\mathcal{M}(X')$. This high-level idea serves as the principle for determining the confidence in our concluded lower bound.

We further introduce the following definition that serves as our basic model for reasoning on the privacy audit problem.

\begin{definition}[$f$-DP Channel]\label{def:f_dp_channel}
    Let $P,Q$ satisfy $T_{P,Q}=f$ and be constructed according to Definition \ref{def:tight_distribution_pair}. Under our audit framework formulation, an $f$-DP channel is framework $(1, \mathcal{M}_f,\mathbf{H}_f, \mathbf{D}_f, 1)$ where
    \begin{itemize}
        % \vspace{-0.2cm}
        \item $\mathbf{H}_f$ is the identity function
        % \vspace{-0.2cm}
        \item $\mathcal{M}_f:\{0,1\}\rightarrow [0,1]$ output one sample from $P$ if input is 0 and output one sample from $Q$ if input is 1
        % \vspace{-0.2cm}
        \item $\mathbf{D}_f$ is some arbitrary decoder
        % \vspace{-0.2cm}
        \item Filtering is trivial since $n=r=1$
    \end{itemize}
\end{definition}
By construction, guessing the input bit to the $f$-DP
channel is as hard as distinguishing the two distributions $P$ and $Q$ based on a single draw. If algorithm $\mathcal{M}$ is $f$-DP, distinguishing $\mathcal{M}(X)$ from $\mathcal{M}(X')$ is even harder than guessing the input bit to the $f$-DP channel. With such preparation, we are ready to find the optimal decoder $\mathbf{D}^{opt}_f$ such that the framework $(1,\mathcal{M}_f,\mathbf{H}_f, \mathbf{D}_f, 1)$ achieves the lowest bit error probability. The following Lemma says what $\mathbf{D}^{opt}_f$ should be:

\begin{lemma}[Optimal $\mathbf{D}^{opt}_f$  for $f$-DP Channel. Proof in Appendix \ref{sec:proof_optimal_decoder_f_dp_channel}]\label{lem:optimal_decoder_f_dp_channel}
    The optimal decoder $\mathbf{D}^{opt}_f$ for $f$-DP channel shown in Definition \ref{def:f_dp_channel} is as follows:
    $\mathbf{D}^{opt}_f(y)$ always outputs score $\mathbf{D}^{opt}_f(y)[0]=\left|\mathcal{L}_{P/Q}(y)\right|$ where $\mathcal{L}_{P/Q}(y) = \ln(P(y)/Q(y))$ is privacy loss at $y$ (Definition \ref{def:pld}). For guessing the bit,
    \begin{equation}\label{equ:optimal_decoder_f_dp_channel}
    % \large
        \mathbf{D}^{opt}_f(y)[1] = \left\{\begin{array}{ll}
            0 & \textit{if } \frac{P(y)}{Q(y)} \geq 1\\
            1 & \textit{if } \frac{P(y)}{Q(y)} <    1\\
        \end{array}\right.
    \end{equation}
    $\mathbf{D}^{opt}_f$ achieves the \textbf{lowest} single bit error probability:
    \begin{equation}\label{equ:lowest_bit_error_prob}
        \inf_{\mathbf{D}_f}\pr{B,\mathcal{M}_f}{E_i=1} =\frac{\alpha^{opt} +f(\alpha^{opt})}{2}
    \end{equation}
    where $\alpha^{opt}\in(0,1)$ is any number such that $-1$ is one subgradient of the trade-off function $f$ at point $\alpha^{opt}$.
\end{lemma}

By just inspecting the shape of the trade-off function, it is easy to see that point $\alpha^{opt}$ is simply the fixed point such that 
$$
\alpha^{opt}=f(\alpha^{opt})
$$ 
for all of the trade-off functions shown in Figure \ref{fig:dif_PQ} (all of them are symmetric in the sense that $f=f^{-1}$).

% \vspace{\MargBtParag}
\notbf{Intuition.} One notable fact about the decision rule shown in Equation \eqref{equ:optimal_decoder_f_dp_channel} is that it is equivalent to the well-known \textit{maximum a posterior} decision rule that minimizes the error probability given a uniform prior. Specifically, given $y$ output by $\mathcal{M}_f$, the rule shown in Equation \eqref{equ:optimal_decoder_f_dp_channel} ensures the choice corresponds to maximum posterior probability. Specifically, by Bayes' Theorem, for framework $(1,\mathcal{M}_f,\mathbf{H}_f, \mathbf{D}_f, 1)$, we have (random bit vector $B$ is one-dimensional in this case)

\begin{large}
\begin{equation}\label{equ:posterior_bit_prob_f_dp_channel}
    B|y = \left\{\begin{array}{ll}
        0 & w.p. \quad \frac{P(y)/Q(y)}{1+ P(y)/Q(y)}\\
        1 & w.p. \quad \frac{1}{1+ P(y)/Q(y)}\\
    \end{array}\right.
\end{equation}
\end{large}

And it is obvious that ${\mathbf{D}^{opt}_f}$ selects the most probable choice among $\{0,1\}$ as its guessed bit. This leads to the following Lemma.
\sheperdrevisionStart
\begin{lemma}[Error probability, proof in Appendix \ref{sec:proof_error_prob}]\label{lem:error_prob}
Conditioned on $y$, $B$ is just a Bernoulli random variable, and for the guessed bit made by ${\mathbf{D}^{opt}_f}$, the bit error is

\begin{large}
\begin{equation}\label{equ:posterior_best_bit_error}
    E|y = \left\{\begin{array}{ll}
        0 & w.p. \quad 1 - \frac{1}{1+ \exp(\left|\mathcal{L}_{P/Q}(y)\right|)}\\
        1 & w.p. \quad \frac{1}{1+ \exp(\left|\mathcal{L}_{P/Q}(y)\right|)}\\
    \end{array}\right.
\end{equation}
\end{large}
\end{lemma}
\sheperdrevisionEnd

Now we are ready to introduce our main theorem for our privacy audit framework.

\begin{theorem}[Transmission dominated by Independent $f$-DP Channels. Proof in Appendix \ref{sec:proof_trans_dominated_by_ind_f_dp_channel}]\label{thm:trans_dominated_by_ind_f_dp_channel}
    If $\mathcal{M}$ is $f$-DP, then for all possible $\mathcal{M}, \mathbf{H}, \mathbf{D}$
    $$
    (n,  \mathcal{M}, \mathbf{H}, \mathbf{D}, r) \preceq (n, \mathcal{M}', \mathbf{H}', \mathbf{D}', r)
    $$
    where, $\mathcal{M}', \mathbf{H}', \mathbf{D}'$ is described as follows:    
    \begin{itemize}
        % \vspace{-0.1cm}
        \item $\mathbf{H}'(b) = X_{in} = b$ is the identity function
        
        % \vspace{-0.2cm}
        \item $\mathcal{M}'$ is $n$ parallel independent runs of $\mathcal{M}_f$ as defined in Definition \ref{def:f_dp_channel}, i.e., $\mathcal{M}'(b) = y = (y_1, y_2, \ldots, y_n)$ where $y_i$ is sampled from $P$ if $b_i=0$ and sampled from $Q$ if $b_i=1$.
        
        % \vspace{-0.2cm}
        \item $\mathbf{D}'$ is $n$ parallel independent runs of $\mathbf{D}^{opt}_f$ as defined in Lemma \ref{lem:optimal_decoder_f_dp_channel}, i.e., $\mathbf{D}'(y) = ((s_1^f, a_1^f), (s_2^f, a_2^f), \ldots, (s_n^f, a_n^f))$ where $s_i = \left|\mathcal{L}_{P/Q}(y_i)\right|$ and $a_i$ is guessed according to Equation \eqref{equ:optimal_decoder_f_dp_channel}.
    \end{itemize}
\end{theorem}

\notbf{Proof sketch.} To prove the above theorem, we first start at $(n, \mathcal{M}, \mathbf{H}, \mathbf{D}, r)$, making a series of comparisons between the two audit frameworks (where one is dominated by the other), and we finally arrive at $(n, \mathcal{M}', \mathbf{H}', \mathbf{D}', r)$. The above theorem says that under the same privacy characterization for the targeted private algorithm, framework $(n, \mathcal{M}', \mathbf{H}', \mathbf{D}', r)$ is the best one can do in terms of making bit errors. The best is achieved by transmitting each bit independently through an $f$-DP channel. 

The above theorem bypasses the difficulties of possible dependencies among guesses and allows us to simplify the problem to the independent case. Technically, to get rid of the dependency issue, \cite{steinke2023privacy} has to add a bad $\mathcal{O}(n)$ factor that deteriorates the final results. In contrast, we are not limited by $\mathcal{O}(n)$, and this is our first key step to achieve tight audit results.

For some $u$ we observe, the remaining work is to compute the tail probability (which determines the confidence level)
\begin{equation}\label{equ:final_tail_prob}
% \large
     \pr{B,\mathcal{M}'}{\mathbf{event}_u}
\end{equation}
for some realization of $(n, \mathcal{M}',\mathbf{H}', \mathbf{D}', r)$ mentioned in Theorem \ref{thm:trans_dominated_by_ind_f_dp_channel} so that we can conclude a privacy lower bound. Theorem \ref{thm:trans_dominated_by_ind_f_dp_channel} guarantees that any possible value of the above tail probability is an upper bound for $\pr{B,\mathcal{M}}{\mathbf{event}_u}$ of the original audit framework $(n,\mathcal{M}, \mathbf{H}, \mathbf{D}, r)$.

% \vspace{\MargBtParag}
\notbf{Discussion on the decoder $\mathbf{D}$.} In our analysis, we quantify how the optimal decoder $\mathbf{D}^{opt}_f$ behaves. The optimal decoder $\mathbf{D}^{opt}_f$ is the one that minimizes the bit error probability for the $f$-DP channel. This is used to compute the best one can do in terms of making bit errors. In practice, the decoder may not be optimal, i.e., the output score is not informative, which makes the guesses bad (e.g., the guess is close to random guessing). Then, no audit method can magically still give tight audit results.
% Then any privacy audit method based on bad guesses will see a gap between the upper bound and the audited lower bound. Xiang et al. \cite{xiang2023privacy} reasoned about this issue by a markov chain 
This issue is not our focus and may be mitigated by the advancements in membership inference attacks.

% \vspace{\smgbs}
\subsection{Compute Tail Bound}\label{sec:privacy_audit_tail_bound}
\vspace{-0.2cm}
\notbf{Problem reduction.} Based on Theorem \ref{thm:trans_dominated_by_ind_f_dp_channel}, we simplify the problem of computing the tail bound as follows. 

We have $n$ independent $f$-DP channels \BestfChannel, where each input bit $b_i$ is sampled from $\mathbf{Bernoulli}(1/2)$, and finally $r$ guessed bits will be released based on the scores output by each decoder of each channel. The filtering obeys Equation \eqref{equ:filter_condition}.

The high-level idea of our analysis is Bayesian, i.e., conditioned on the output of each $\mathcal{M}_f$, we analyze and understand the filtering behavior. Specifically, from a posterior view,
\begin{large}
\begin{equation}\label{equ:posterior_bit_prob_f_dp_channel_each_coor}
    B_i|y_i = \left\{\begin{array}{ll}
        0 & w.p. \quad \frac{P(y_i)/Q(y_i)}{1+ P(y_i)/Q(y_i)}\\
        1 & w.p. \quad \frac{1}{1+ P(y_i)/Q(y_i)}\\
    \end{array}\right.
\end{equation}
\end{large}
by Equation \eqref{equ:posterior_bit_prob_f_dp_channel}; making guesses according to $\mathbf{D}^{opt}_f$, we have
$$
\pr{}{E_i=1|y_i} = \frac{1}{1+ \exp(\left|\mathcal{L}_{P/Q}(y_i)\right|)}
$$
i.e.,
$$
E_i|y_i \sim \mathbf{Bernoulli}( \frac{1}{1+ \exp(\left|\mathcal{L}_{P/Q}(y_i)\right|)})
$$

Then, the problem is reduced to the following: given $n$ independent samples $\{(s_i^f, e_i^f): i\in[n]\}$ where $(s_i^f, e_i^f)$ is generated by
\begin{enumerate}
    % \vspace{-0.2cm}
    \item First sample $y_i$ from $\frac{P+Q}{2}$
    % \vspace{-0.3cm}
    \item Then compute $s_i^f=\left|\mathcal{L}_{P/Q}(y_i)\right|$
    % \vspace{-0.3cm}
    \item Then sample $e_i^f$ from $\mathbf{Bernoulli}( \frac{1}{1+ \exp(\left|\mathcal{L}_{P/Q}(y_i)\right|)})$
    % \vspace{-0.3cm}
\end{enumerate}
And then we need to release $r$ error-indicator bits from $\{e_i^f: i\in[n]\}$ obeying Equation \eqref{equ:filter_condition} by their corresponding scores. 
\RevisionChangeStart
Writing
\begin{equation}\label{equ:released_error_count}
    T \;:=\; \mathbf{SUM}\big(\{E_i: \hat{B}_i\neq\bot\}\big)
\end{equation}
for the number of mistakes among the released guesses, Theorem
\ref{thm:trans_dominated_by_ind_f_dp_channel} guarantees that any upper bound on
$\pr{}{T\leq u}$ in this experiment is also an upper bound for the original audit:
\begin{equation}\label{equ:valid_tail_prob_upper_bound}
    \pr{B,\mathcal{M}}{\mathbf{event}_u} \;\leq\; \pr{}{T\leq u} .
\end{equation}
\RevisionChangeEnd
There is one last preparation before we can tractably compute a valid realization of the right-hand side of Equation \eqref{equ:valid_tail_prob_upper_bound}. If for some $y_i,y_j\in(0,1), y_i\neq y_j$ we have $s_i^f=s_i^f$, we will enforce a tie-breaking rule such that $e_i^f$ has a higher priority to be released if
\begin{equation}\label{equ:tie_breaking_rule}
    y_i > y_j, \text{ given }s_i^f=s_j^f
\end{equation}
Note that, no matter what tie-breaking rule we choose, the final result for the right-hand side of Equation \eqref{equ:valid_tail_prob_upper_bound} is still a valid upper bound for $\pr{B,\mathcal{M}}{\mathbf{event}_u}$, this is because we never violate the filtering condition in Equation \eqref{equ:filter_condition}, and Theorem \ref{thm:trans_dominated_by_ind_f_dp_channel} guarantees its validity. We set the above tie-breaking rule just to make the problem analytically solvable. The rest of this section computes the right-hand side.

\RevisionChangeStart
\notbf{Score quantiles.} Let $F_S$ be the c.d.f.\ of the score, so that
$F_S(s)$ is the fraction of outputs scoring below $s$; concretely
\begin{equation}\label{equ:score_cdf}
\begin{aligned}
    & F_S(s)=\int_{R_s} f_Y(y)\,dy\\
    & R_s=\Big\{y' : \left|\mathcal{L}_{P/Q}(y')\right|<s\Big\}
\end{aligned}
\end{equation}
together with the tie-breaking rule of Equation \eqref{equ:tie_breaking_rule}. Writing
$U_i:=F_S(S_i)$, the $U_i$ are i.i.d.\ uniform on $[0,1]$ (This is well known, as any continuous random variable will be transformed to be uniform on $[0,1]$ by its c.d.f.).We first introduce the following Lemma.

\begin{lemma}[What $\gamma$ and $\bar{\gamma}$ are, Proof in Appendix \ref{sec:proof_markov_chain}]\label{lem:gbar}
define
$$
  g(x) = \frac{1}{1+e^x},\quad\gamma(w) := g\big(F_S^{-1}(w)\big)
$$
then, $\gamma(w)$ is the error probability of a guess at score quantile $w$, and 
$$
  \bar{\gamma}(w) := \frac{1}{1-w}\int_w^1 \gamma(w')\,dw'
$$
as the mean error rate \emph{above} quantile  $w$ (marginalize above $w$).
For a single channel, and for every $w\in[0,1)$,
\begin{equation}\label{eq:gbar-meaning}
\begin{aligned}
    \pr{}{E=1 , U=w}=\gamma(w)\\
    \pr{}{E=1 | U>w}=\bar{\gamma}(w)
\end{aligned}
\end{equation}

\end{lemma}
This technical Lemma allows us to work in the world of score quantiles and will make analysis easier.
\RevisionChangeEnd

\notbf{Order statistics.} We then find that the problem is essentially and closely related to \textit{order statistics}, and we can reach tight audit results based on this modeling. Note that this implication is not discovered by previous work \cite{steinke2023privacy,xiang2025privacy,mahloujifar2024auditing}.

Specifically, we introduce the following definition.

\begin{definition}\label{def:k-th_order_stats_def}
Suppose we have $n$ i.i.d.\ samples from $S$, denoted as $S_1, S_2, \cdots, S_n$. The
$k$-th order statistic of these samples is defined as the $k$-th smallest one among them,
denoted as $S_{(k)}$. And we have
\begin{equation}\label{equ:order_statistic}
\begin{aligned}
    S_{(1)} =& \min\{S_1, S_2, \ldots, S_n\}\\
    S_{(k)} =& \text{ the $k$-th smallest in } \{S_1, S_2, \cdots, S_n\}\\
    S_{(n)} =& \max\{S_1, S_2, \ldots, S_n\}
\end{aligned}
\end{equation}
\end{definition}

\RevisionChangeStart
\notbf{Order statistics are a Markov chain.} The filtering in our protocol is a rank operation, so we need
the joint behaviour of the ordered scores. One classical fact supplies all of it.

\begin{lemma}[Markov property and block form. Proof in Appendix \ref{sec:proof_markov_chain}]\label{lem:markov_chain}
    Let $Z_1,\dots,Z_n$ be i.i.d.\ with density $g$ and c.d.f.\ $G$. Then
    $(Z_{(1)},\dots,Z_{(n)})$ is a Markov chain: conditioning on $(Z_{(1)},\dots,Z_{(i)})$
    is the same as conditioning on $Z_{(i)}$ alone. Moreover, conditioned on $Z_{(i)}=z$,
     $(Z_{(i+1)},\dots,Z_{(n)})$ is independent of $(Z_{(1)},\dots,Z_{(i-1)})$ and
    is distributed as the order statistics of $n-i$ i.i.d.\ draws from $g$ truncated to
    $(z,\infty)$.
\end{lemma}

This lemma say that: \emph{everything above $Z_{(i)}$ forgets everything below it, and is simply $n-i$
fresh i.i.d.\ draws conditioned on greater than $Z_{(i)}$.} In the following, we will see that such conditional-i.i.d. property admits a concentration argument that will be used to compute the tail bound. 

% ---------------------------------------------------------------------
\notbf{The cut.} We apply Lemma \ref{lem:markov_chain} at $i=n-r$, that is, we condition on
the highest-scoring guess that is \textbf{discarded}. We call it the \emph{cut} and write
\begin{equation}\label{equ:cut_quantile}
    W:=F_Y(Y_{(n-r)})
\end{equation}
for its score quantile.
% with the convention $W\equiv 0$ when $r=n$ and nothing is discarded.

\begin{lemma}[Released error bits are i.i.d.\ given the cut. Proof in Appendix \ref{sec:proof_cut}]\label{lem:cut}
    Conditioned on the cut, the $r$ released error indicators are i.i.d.\
    $\mathbf{Bernoulli}\big(\bar{\gamma}(W)\big)$. Equivalently,
    \begin{equation}\label{equ:cut_law}
    \begin{aligned}
        W&\sim\mathbf{Beta}(n-r,\;r+1)\\
        T \,\big|\, W \;&\sim\; \mathbf{Binomial}\big(r,\;\bar{\gamma}(W)\big) 
    \end{aligned}
    \end{equation}
\end{lemma}

With such preparation, the following theorem computes the tail bound by averaging Lemma \ref{lem:cut} over the cut.

\begin{theorem}[Tail Probability for Privacy Audit. Proof in Appendix \ref{sec:proof_best_tail_bound}]\label{thm:best_tail_bound}
    Given the setup in Theorem \ref{thm:trans_dominated_by_ind_f_dp_channel} and the
    tie-breaking rule in Equation \eqref{equ:tie_breaking_rule}, for every
    $u\in\{0,1,\dots,r\}$,
    \begin{equation}\label{equ:best_tail_bound}
    \begin{aligned}
        &\pr{B,\mathcal{M}}{\mathbf{event}_u}
        \;\leq\;
        \E{W}{\;\pr{}{\mathbf{Binomial}\big(r,\bar{\gamma}(W)\big)\leq u}\;}\\
    \end{aligned}
    \end{equation}
    where $W\sim\mathbf{Beta}(n-r,\,r+1)$ and the right-hand side is \emph{exactly} the tail probability of the released error
    count in the framework $(n,\mathcal{M}',\mathbf{H}',\mathbf{D}',r)$. By Lemma \ref{lem:cut},
    \begin{equation}\label{equ:compute_p_k}
    \begin{aligned}
        &\E{W}{\pr{}{\mathbf{Binomial}\big(r,\bar{\gamma}(W)\big)\leq u}}\\
        =& \int_0^1 \pi_{n,r}(w)\,\mathbf{BinCDF}\big(u;\,r,\,\bar{\gamma}(w)\big)\,dw
    \end{aligned}
    \end{equation}
    where $\mathbf{BinCDF}(u;m,p)=\pr{}{\mathbf{Binomial}(m,p)\leq u}$ and
    \begin{equation}\label{equ:cut_density}
        \pi_{n,r}(w)=\frac{n!}{(n-r-1)!\,r!}w^{\,n-r-1}(1-w)^{r}
    \end{equation}
    is the density of the cut. 
\end{theorem}
\RevisionChangeEnd

\notbf{\underline{Significance of using order statistics}.} Modelling the filtering action
(inspired by the abstaining action in \cite{steinke2023privacy}) through order statistics
captures the implication behind it. Intuitively, suppose we release only $1$ guess,
corresponding to the highest score out of $10^{100}$ guesses; we should expect that single
released guess to be almost perfectly accurate. Previous work treats such a guess as no
different from a randomly chosen one, leading to a loose modeling and hence a loose
lower bound. Lemma \ref{lem:cut} makes the adjustment precise: the released guesses are
$r$ draws from the channel \emph{conditioned on being greater than the cut}, so their error rate is
$\bar{\gamma}$ evaluated above the cut rather than the unconditional counterpart.

Now we have all the information to implement the audit; the details of implementation are
included in the Appendix \ref{sec:audit_method} for reference. 
% Our implementation is in an
% anonymous link ``{\it anonymous.4open.science/r/TPAiORSP}''.

\begin{figure*}[!ht] 
    \centering

    \subfloat[$0.4$-GDP]
    {
    \includegraphics[width=.33\linewidth]{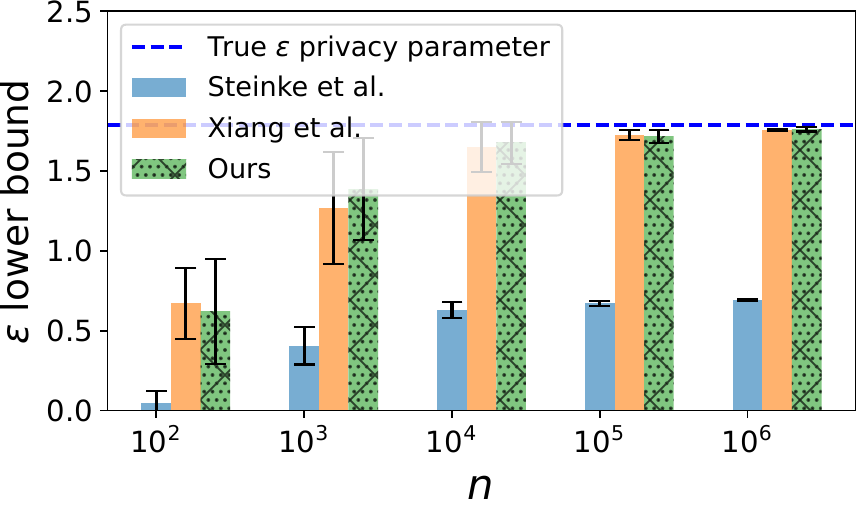}
    }
    \subfloat[$0.8$-GDP]
    {
    \includegraphics[width=.33\linewidth]{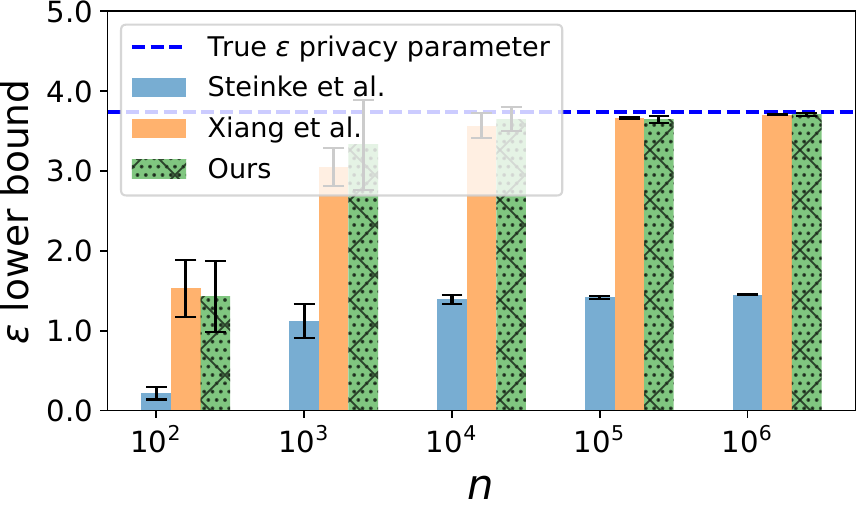}
    }
    \subfloat[$1.6$-GDP]
    {
    \includegraphics[width=.33\linewidth]{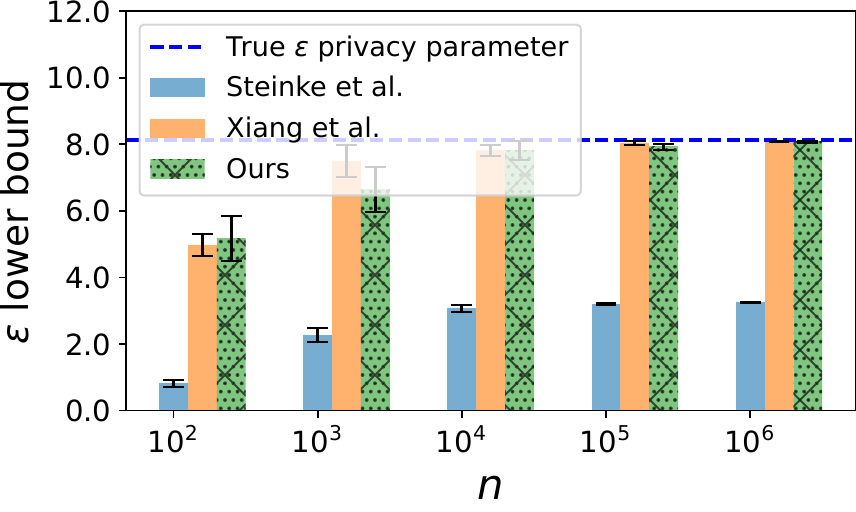}
    }
    % \subfloat[$\mu=3.2$]
    % {
    % \includegraphics[width=.33\linewidth]{figures_tables/n_vs_eps_lb_gdp3.2.pdf}
    % }
    \vspace{\fmgftoc}
    \caption{
    \RevisionChangeStart
    Audit result comparison on Gaussian mechanism in the ``special case''. As guesses are independent, Xiang et al.'s \cite{xiang2025privacy} method is valid in this case. For ours and Steinke et al.'s \cite{steinke2023privacy} method, we set the same $r = n/5$ to ensure a fair comparison. $\delta=10^{-5}$.
    \RevisionChangeEnd
    }
    \label{fig:gdp_audit} 
    \vspace{\fmgctom}
\end{figure*}

\begin{figure*}[!ht] 
    \centering

    \subfloat[$\operatorname{Lap}(0,1)\text{ V.S. }\operatorname{Lap}(0.4,1)$]
    {
    \includegraphics[width=.33\linewidth]{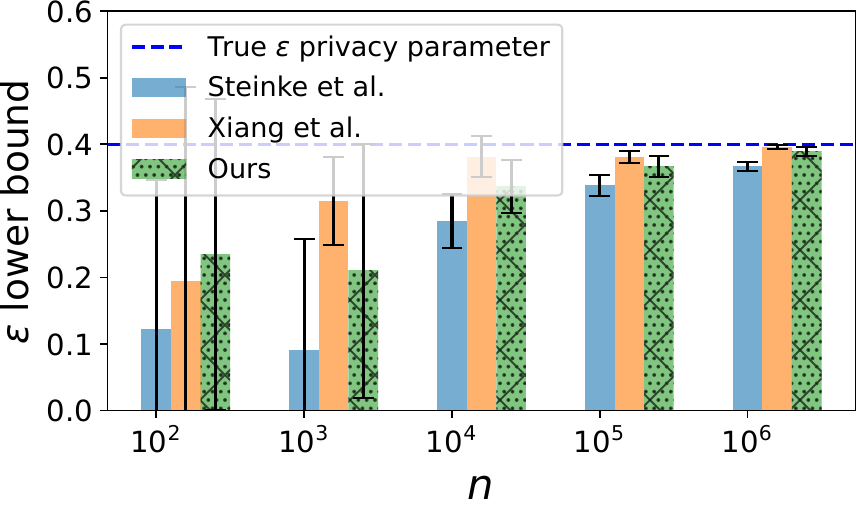}
    }
    \subfloat[$\operatorname{Lap}(0,1) \text{ V.S. }\operatorname{Lap}(0.8,1)$]
    {
    \includegraphics[width=.33\linewidth]{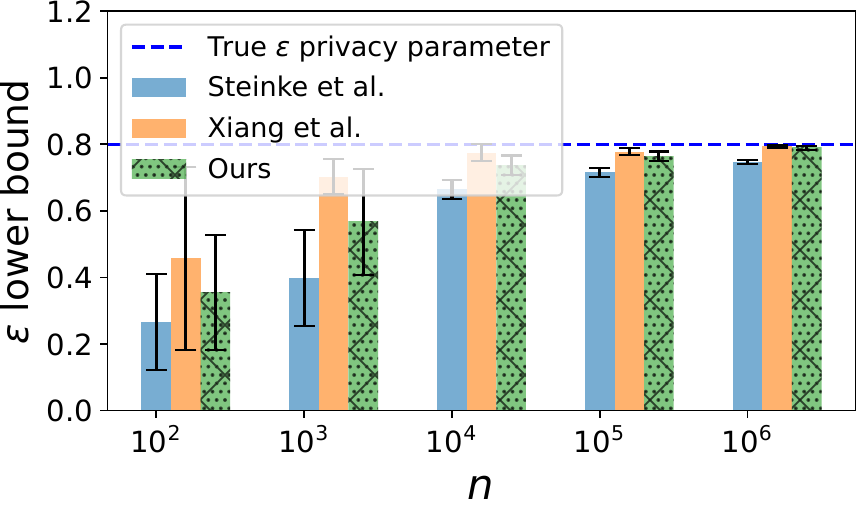}
    }
    \subfloat[$\operatorname{Lap}(0,1) \text{ V.S. }\operatorname{Lap}(1.6,1)$]
    {
    \includegraphics[width=.33\linewidth]{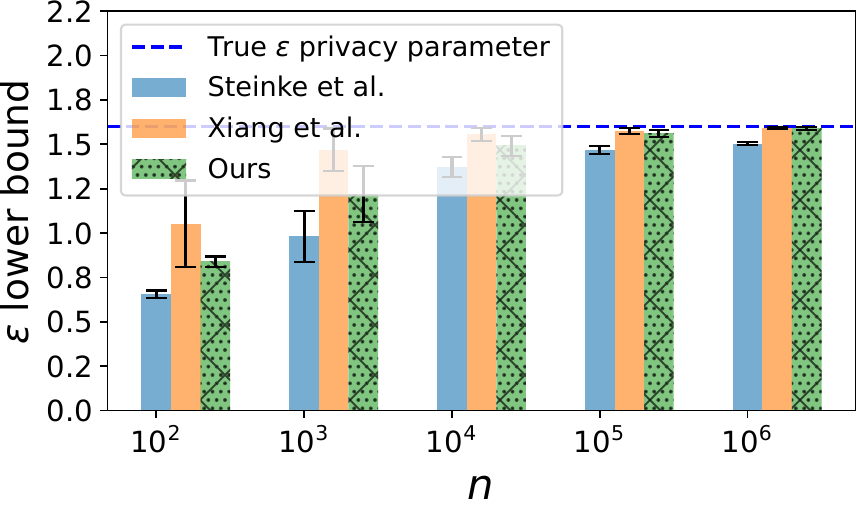}
    }
    % \subfloat[$\mu=3.2$]
    % {
    % \includegraphics[width=.33\linewidth]{figures_tables/n_vs_eps_lb_gdp3.2.pdf}
    % }
    % \vspace{\fmgftoc}
    \caption{
    \RevisionChangeStart
    Audit result comparison on Laplace mechanism in the ``special case''. As guesses are independent, Xiang et al.'s \cite{xiang2025privacy} method is valid in this case. For ours and Steinke et al.'s \cite{steinke2023privacy} method, we set the same $r = n/5$ to ensure a fair comparison. $\delta=10^{-5}$.
    \RevisionChangeEnd
    }
    \label{fig:laplace_audit} 
    % \vspace{\fmsgctom}
\end{figure*}

\begin{figure*}[!ht] 
    \centering

    \subfloat[$0.5$-GDP]
    {
    \includegraphics[width=.33\linewidth]{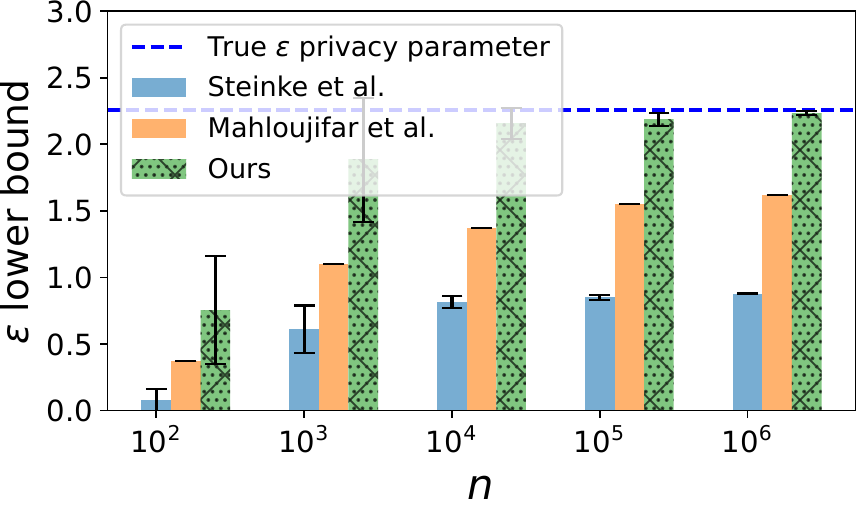}
    }
    \subfloat[$1$-GDP]
    {
    \includegraphics[width=.33\linewidth]{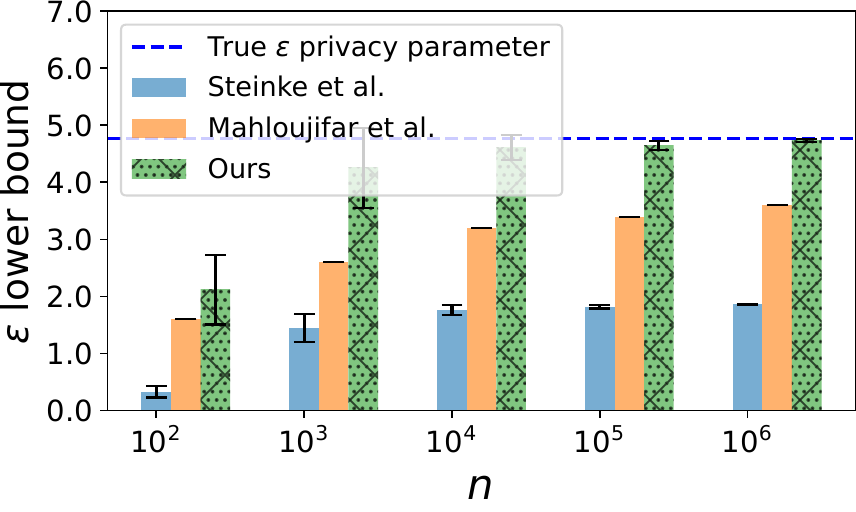}
    }
    \subfloat[$2$-GDP]
    {
    \includegraphics[width=.33\linewidth]{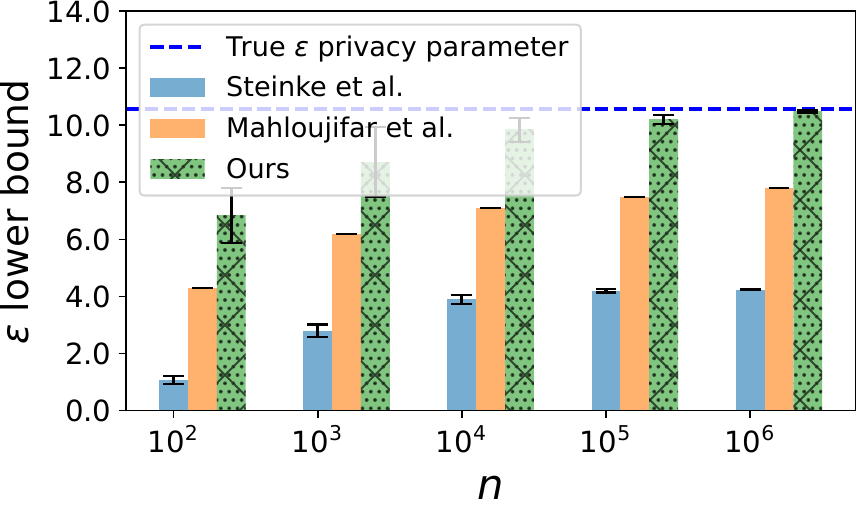}
    }

    \vspace{\fmgftoc}
    \caption{
    \RevisionChangeStart
    Audit result comparison on Gaussian mechanism in the ``general case''. Since only a subset of guesses are released, which introduces dependencies, Xiang et al.'s \cite{xiang2025privacy} method is not valid in this case. We compare our method with Steinke et al.'s \cite{steinke2023privacy} and Mahloujifar et al.'s \cite{mahloujifar2024auditing} methods. We set the same $r = n/5$ for ours and Steinke et al.'s. We use the best reported result of Mahloujifar et al.'s \cite{mahloujifar2024auditing}. $\delta=10^{-5}$.
    \RevisionChangeEnd
    }
    \label{fig:gdp_audit_meta_steinke} 
    \vspace{\fmgctom}
\end{figure*}

\section{Evaluation}\label{sec:evaluation}
\vspace{-0.2cm}
In the experiment, we first compare with previous work \cite{steinke2023privacy,xiang2025privacy,mahloujifar2024auditing} on privacy audit in one run. Then, we give several ablation studies on our framework. For reporting the audited privacy lower bound, we use $(\varepsilon,\delta)$-DP form, as it is the most widely used form in previous work \cite{steinke2023privacy,xiang2025privacy,mahloujifar2024auditing}. Throughout all experiments, we use a confidence level of $95\%$. The error bar is obtained by running each experiment $5$ times.

% \vspace{\smgbs}
\subsection{Comparison: Special Case}\label{sec:comparison_special}
\vspace{-0.2cm}
``Special case'' means that the guesses are independent, i.e., each pair $(b_i,\hat{b}_i)$ is independent from $(b_j,\hat{b}_j)$ for $i\neq j$. Obviously, privacy audit by many independent runs belongs to this case, as each run independently generates one $(b_i,\hat{b}_i)$ pair. In this special case, all guessed bits must be released for Xiang et al.'s \cite{xiang2025privacy} method, because if there is any non-trivial filtering, arbitrary dependencies between $(b_i,\hat{b}_i)$ will be introduced.

\notbf{Audit Gaussian mechanism.} For a concrete example, we can consider the Gaussian mechanism  where each bit is perturbed by independent Gaussian noise:
\begin{equation}\label{equ:gm}
    \mathcal{M}(b)=b+\mathcal{N}(0,\sigma^2\mathbb{I}^n)
\end{equation} 
where each coordinate of $b$ is i.i.d. from $\mathbf{Bernoulli}(1/2)$. The guesses are conducted coordinate-wise. And we can see that such a mechanism is $1/\sigma$-GDP with respect to the flip of each bit. The guess criterion is simple, \textit{we guess $\hat{b}_i=0$ if $(\mathcal{M}(b))_i\leq 1/2$, otherwise $\hat{b}_i=1$}. 

Results are presented in Figure \ref{fig:gdp_audit}. We compare our method with Steinke et al.'s \cite{steinke2023privacy} and Xiang et al.'s \cite{xiang2025privacy} methods. Even if our method only guesses over a subset of bits, the results show that our method is almost as good as Xiang et al.'s \cite{xiang2025privacy} method under all settings. Our audit method closes the gap between the lower bound and the privacy upper bound, given $n$ is large enough. Our method also outperforms Steinke et al.'s \cite{steinke2023privacy} method in all cases.

\notbf{Audit Laplace mechanism in the ``special case''.} The setting in this case is similar to the Gaussian mechanism, except that we use Laplace noise instead of Gaussian noise. Specifically, each bit is perturbed by independent Laplace noise:
\begin{equation}\label{equ:lm}
    \mathcal{M}(b)=b+\operatorname{Lap}(0,c)^n
\end{equation} 
And the trade-off function is 
% \begin{normal}
$$
T_{\operatorname{Lap}(0,1), \operatorname{Lap}(1/c, 1)}
$$  
% \end{normal}
(Equation \ref{equ:laplace_dp}). Laplace mechanism satisfies pure differential privacy, i.e., $(\varepsilon,\delta=0)$-DP. \textit{The guessing strategy is the \textit{same} as the above Gaussian mechanism}.

Results are presented in Figure \ref{fig:laplace_audit}. We compare our method with Steinke et al.'s \cite{steinke2023privacy} and Xiang et al.'s \cite{xiang2025privacy} methods. Again, our method is almost as good as Xiang et al.'s \cite{xiang2025privacy} method under all settings. We can also see that \cite{steinke2023privacy}'s method is also tight in this experiment, and this aligns with the observation in \cite{steinke2023privacy} that their method is only tight for pure differential privacy.

\begin{figure}[!ht] 
    \centering

    % \subfloat[$\mu=0.4$]
    {
    \includegraphics[width=.9\linewidth]{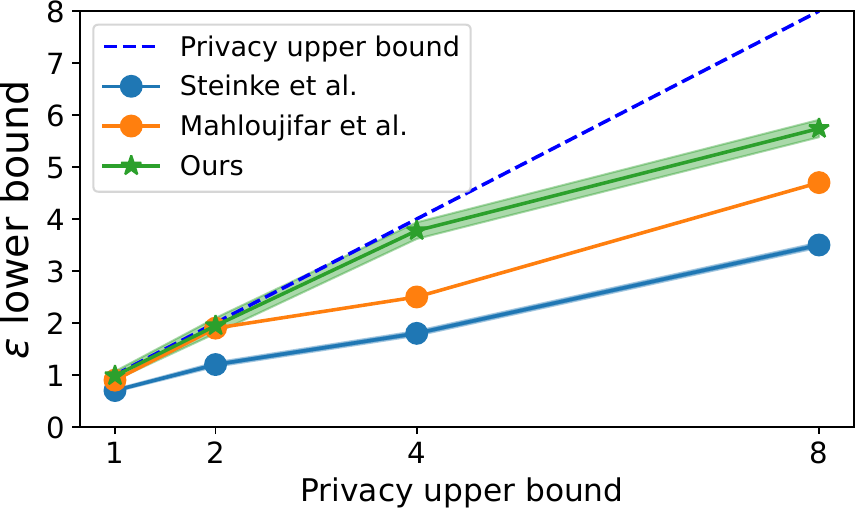}
    }

    \caption{
    \RevisionChangeStart
    Audit result comparison on DP-SGD. We compare our method with Steinke et al.'s \cite{steinke2023privacy} and Mahloujifar et al.'s \cite{mahloujifar2024auditing} methods. We set the same $n=10^{4}, r = n/2$ for ours and Steinke et al.'s to ensure fair comparison. $\delta=10^{-5}$.
    \RevisionChangeEnd
    }
    \label{fig:gdp_audit_dpsgd} 
    % \vspace{\fmgctom}
\end{figure}

\begin{figure}[!ht] 
    \centering

    \subfloat[$\delta=10^{-4}$]
    {
    \includegraphics[width=.9\linewidth]{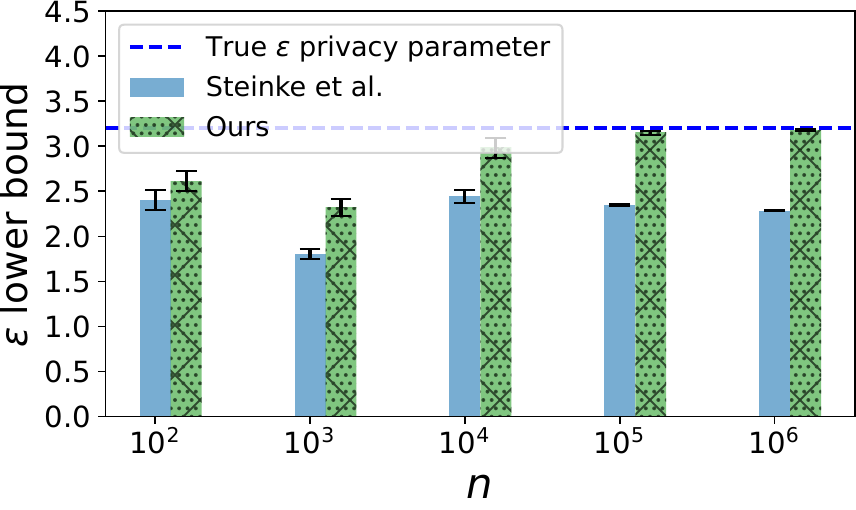}
    }

    \subfloat[$\delta=10^{-2}$]
    {
    \includegraphics[width=.9\linewidth]{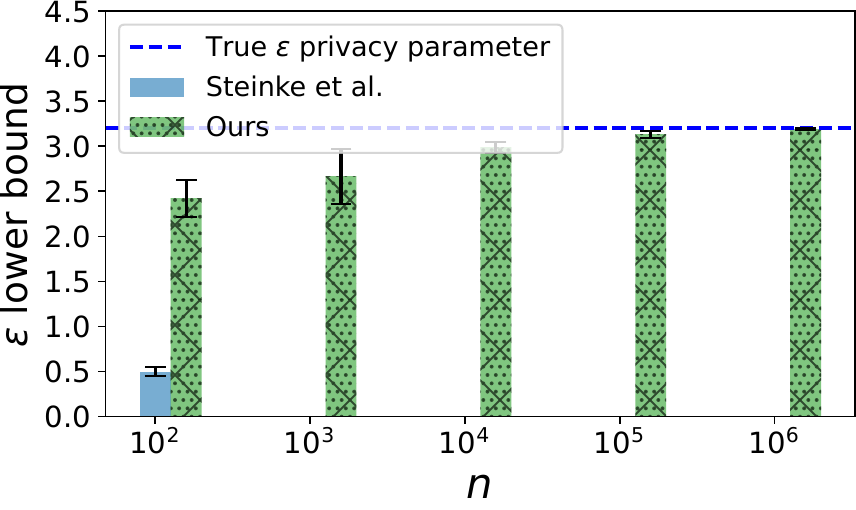}
    }

    \vspace{\fmgftoc}
    \caption{
    \RevisionChangeStart
   Audit comparison on the randomized response mechanism. We compare with Steinke et al.'s \cite{steinke2023privacy} method. We set $\varepsilon=3.2$ and inspect two different $\delta$ values. We set the same $r = n/5$ for ours and Steinke et al.'s. When $\delta=10^{-2}, $\cite{steinke2023privacy} fails to give any meaningful result when $n>10^{2}$.
   \RevisionChangeEnd
   }
    \label{fig:gdp_audit_rr} 
    \vspace{\fmgctom}
\end{figure}

\begin{figure}[!t] 
    \centering
    % \subfloat[$(0.5,0.1)$-DP]
    {
    \includegraphics[width=.95\linewidth]{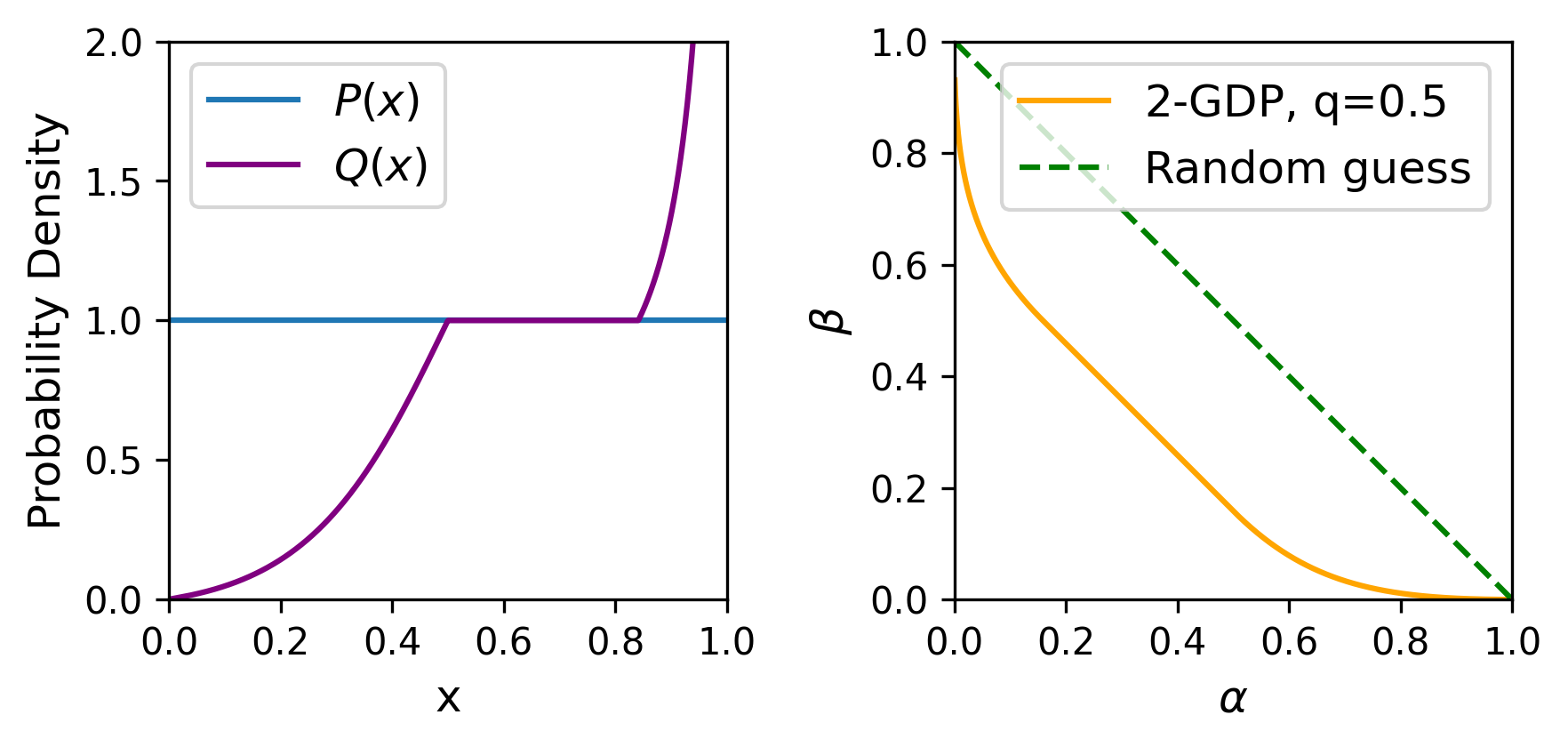}
    }
    \vspace{\fmgftoc}
    \caption{
    The base distribution pairs $P,Q$ for the sub-sampled Gaussian mechanism $1/\sigma$-GDP, $q=0.5$.
    }
    \label{fig:dif_PQ_sgm} 
    \vspace{\fmgctom}
\end{figure}

\begin{figure*}[!ht] 
    \centering

    \subfloat[$0.2$-GDP, $q=0.5$]
    {
    \includegraphics[width=.33\linewidth]{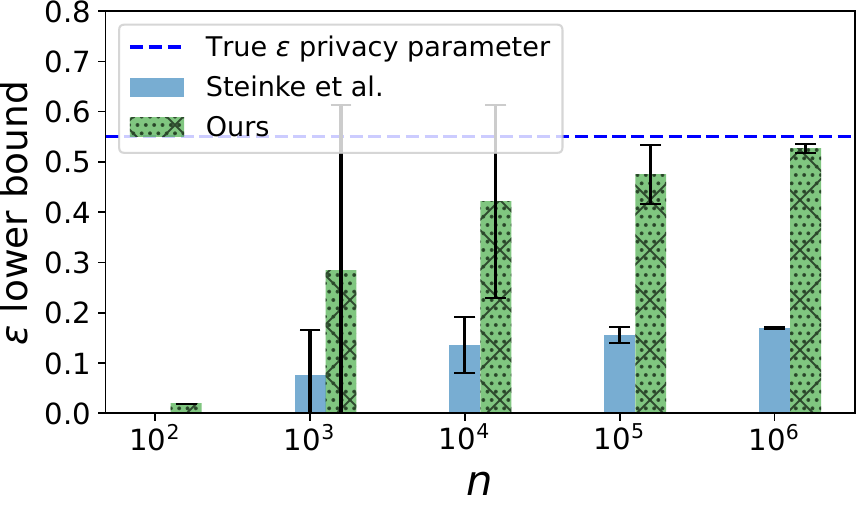}
    }
    \subfloat[$0.4$-GDP, $q=0.5$]
    {
    \includegraphics[width=.33\linewidth]{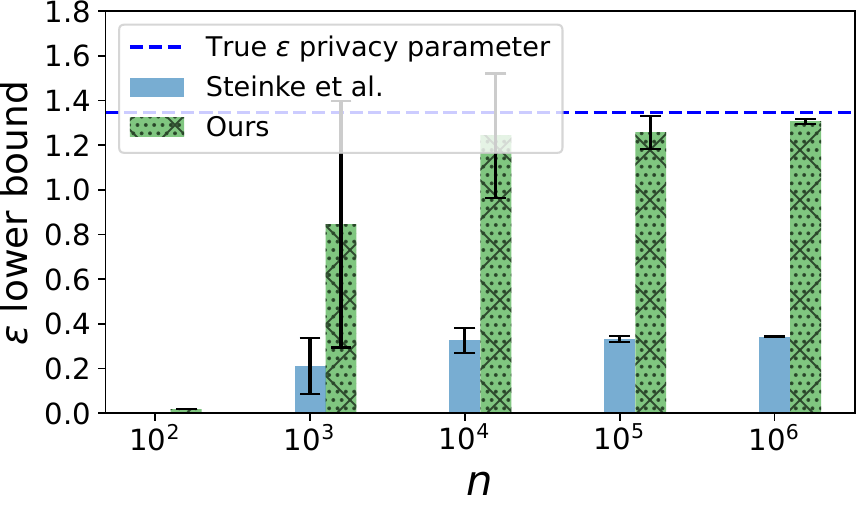}
    }
    \subfloat[$0.8$-GDP, $q=0.5$]
    {
    \includegraphics[width=.33\linewidth]{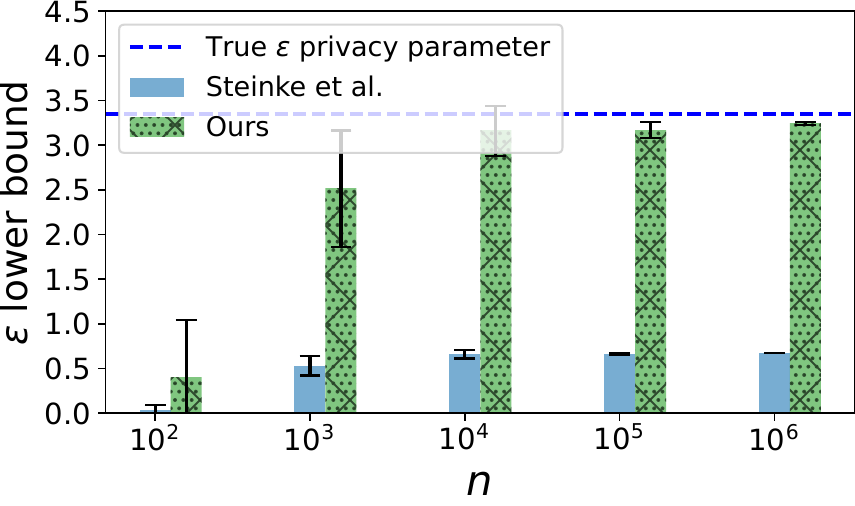}
    }
    \caption{
    \RevisionChangeStart
    Audit the sub-sampled Gaussian mechanism. We compare with Steinke et al.'s \cite{steinke2023privacy}. We set the same $r = n/5$ for ours and Steinke et al.'s. To ensure $\delta$ does not have a significant impact on \cite{steinke2023privacy}'s method, we set it to be very small: $\delta=10^{-7}$.
    \RevisionChangeEnd
    }
    \label{fig:gdp_audit_sgm} 
    % \vspace{\fmgctom}
\end{figure*}

\begin{figure*}[!ht] 
    \centering

    \subfloat[$\delta=10^{-2}$]
    {
    \includegraphics[width=.4\linewidth]{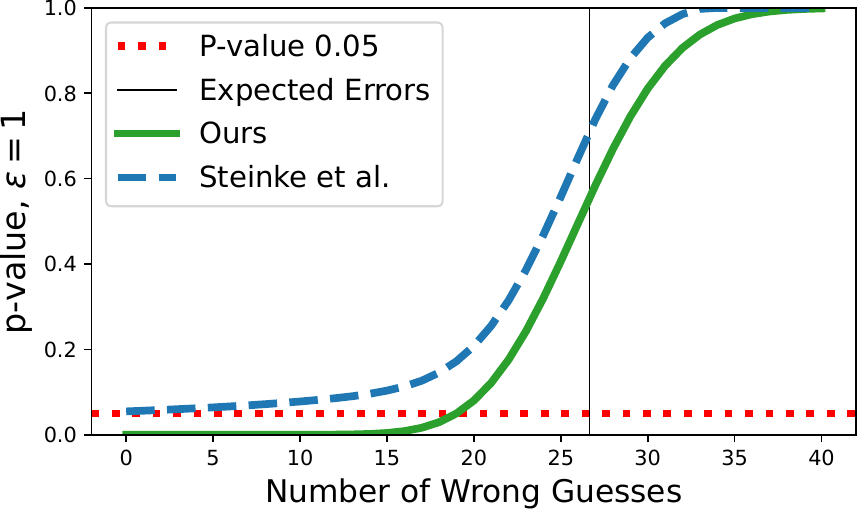}
    }
    \subfloat[$\delta=10^{-1}$]
    {
    \includegraphics[width=.4\linewidth]{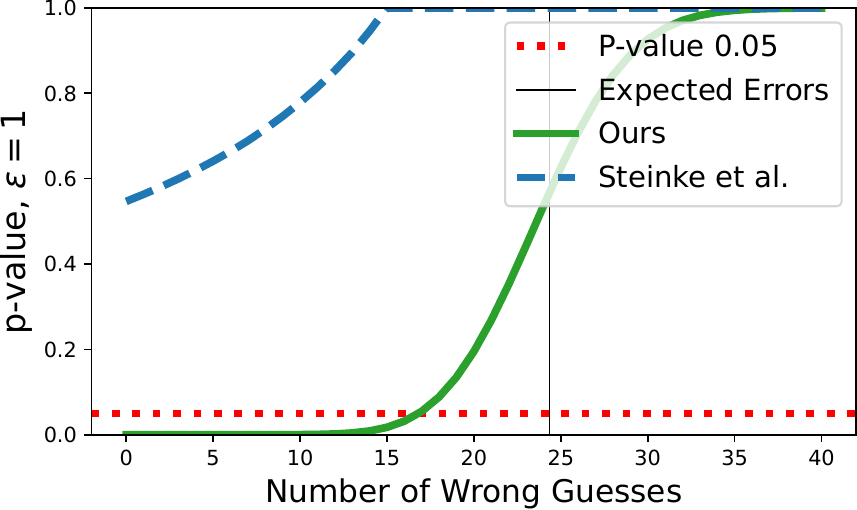}
    }

    \vspace{\fmgftoc}
    \caption{
    \RevisionChangeStart
    P-value comparison between our method and Steinke et al.'s \cite{steinke2023privacy} method. We audit the randomized response mechanism with $\varepsilon=1$ and $\delta=0.01$ or $0.1$. We set the same $n=r = 100$ for both methods. The expected Errors mean that, under each privacy guarantee, the expected bit errors that will be made for $n=100$. 
    \RevisionChangeEnd
    }

    \label{fig:p_value} 
    \vspace{\fmgctom}
\end{figure*}

\begin{figure}[!ht] 
    \centering

    % \subfloat[$\delta=10^{-6}$]
    {
    \includegraphics[width=.9\linewidth]{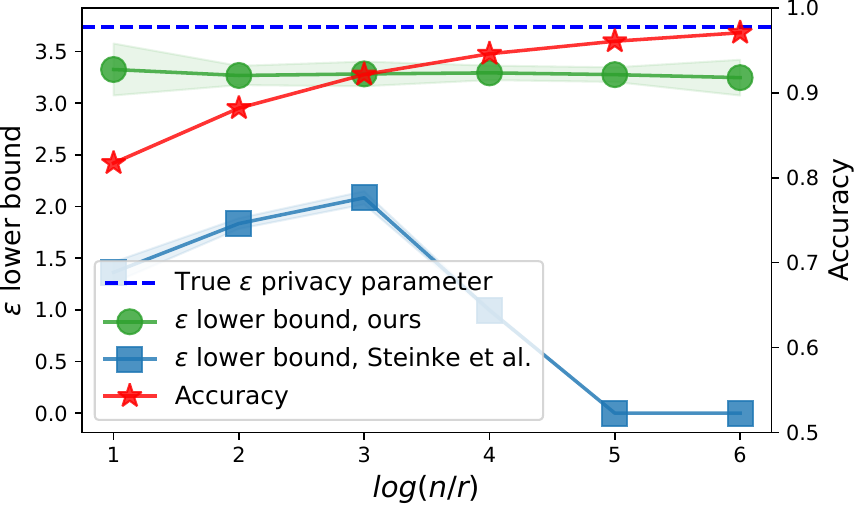}
    }

    % \vspace{\fmgftoc}
    \caption{
    \RevisionChangeStart
    Audit result for Gaussian mechanism with fixed $r=1000$ and varying $n$. We compare with Steinke et al.'s \cite{steinke2023privacy} method. $\delta=10^{-5}$.
    \RevisionChangeEnd
    }
    \label{fig:fix_r_dif_n} 
    % \vspace{\fmgctom}
\end{figure}

\vspace{\smgbs}
\subsection{Comparison: General Case}\label{sec:comparison_general}
\vspace{-0.2cm}
``General case'' means that the guesses are not independent, i.e., there is some non-trivial dependency between $(b_i,\hat{b}_i)$ and $(b_j,\hat{b}_j)$ for some $i\neq j$. For example, if the filtering is non-trivial ($r<n$) and we can only see $r$ guessed bits,  independencies are introduced because of the filtering action: whether some bits are released or not depends on the others. If some input bits in the bit vector $b$ suffer from some correlation in the execution of the private algorithm $\mathcal{M}$, then the guesses are also unavoidably dependent.

This case is what we expect in practice, and the technique of privacy audit in one run is partially motivated by this case. We make comparisons for the ``general case''in the following experiments.

% \vspace{\MargBtParag}
\notbf{Audit Gaussian Mechanism.} We follow the same guessing strategy as that in \cite{steinke2023privacy}: \textit{we choose $r/2$ coordinates having the smallest value to be guessed with 0, and another $r/2$ having the largest value to be guessed with 1}. 

We present results in Figure \ref{fig:gdp_audit_meta_steinke}. We compare our method with Steinke et al.'s \cite{steinke2023privacy} and Mahloujifar et al.'s \cite{mahloujifar2024auditing} method. We can see that for the general case in auditing the Gaussian mechanism, our method outperforms the other methods in all cases. Again, our method closes the gap between the lower bound and the privacy upper bound given $n$ is large enough. \textit{And our method is the first method to reach tight audit result for the general case where $\delta>0$}. In contrast, Steinke et al.'s \cite{steinke2023privacy} method is tight when $\delta=0$ but cannot reach a tight audit result for $\delta>0$ even when all parameter setups are extensively tuned. It is also true for Mahloujifar et al.'s \cite{mahloujifar2024auditing} method, where no tight result can be obtained.

\vspace{\MargBtParag}
\notbf{Audit DP-SGD.} We also audit the DP-SGD algorithm \cite{abadi2016deep} with our method, and compare with Steinke et al.'s \cite{steinke2023privacy} and Mahloujifar et al.'s \cite{mahloujifar2024auditing} method. The experimental setup follows \cite{steinke2023privacy}, and details can be found in Appendix \ref{app:dpsgd}.

The results are presented in Figure \ref{fig:gdp_audit_dpsgd}. We can see that, for auditing some real learning tasks with DP-SGD, our method outperforms the other methods in all setups. Moreover, our method arguably provides a strong lower bound when $\varepsilon=8$, which is a common privacy level used for DP machine learning in practice.

\begin{figure}[!ht] 
    \centering

    \subfloat[$\delta=10^{-6}$]
    {
    \includegraphics[width=.9\linewidth]{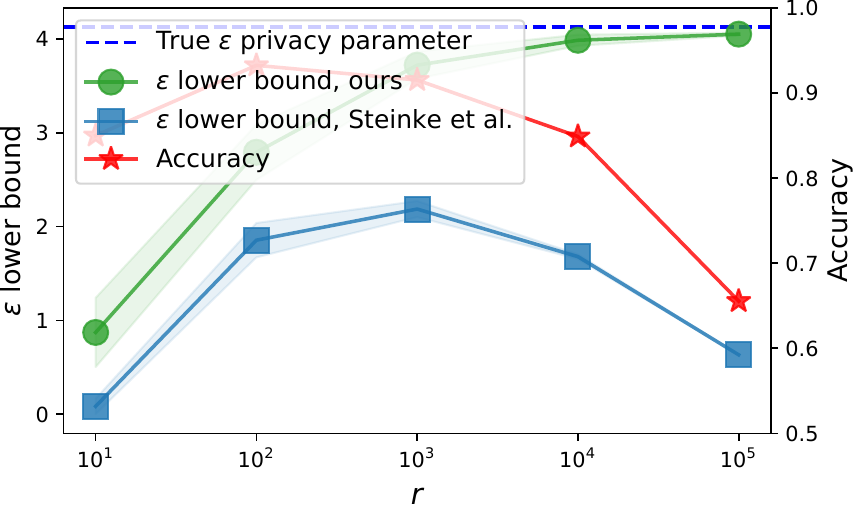}
    }
    
    \subfloat[$\delta=10^{-4}$]
    {
    \includegraphics[width=.9\linewidth]{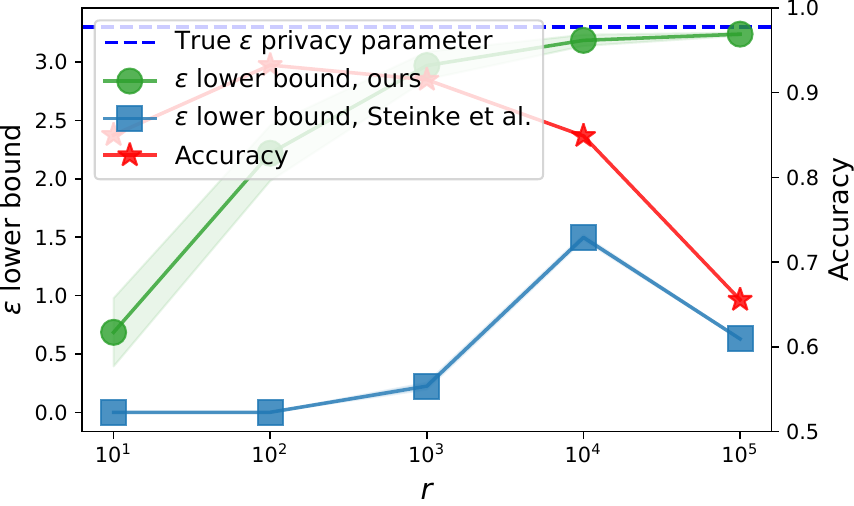}
    }
    \vspace{\fmgftoc}
    \caption{
    \RevisionChangeStart
    Audit result for Gaussian mechanism satisfying $0.8$-GDP. The audit setup is the same as that in Section \ref{sec:comparison_general}. We compare our method with Steinke et al.'s \cite{steinke2023privacy} method. We set the same $n=10^{5}$, and vary $r$. The left vertical axis is the privacy lower bound $\varepsilon$, the right vertical axis is the accuracy $c/r$, where $c$ is the number of correct guesses. 
    \RevisionChangeEnd
    }
    \label{fig:gdp_audit_r_trade_off} 
    \vspace{\fmgctom}
\end{figure}

% \vspace{\MargBtParag}
\notbf{Audit randomized response mechanism under some large $\delta$ value.}  We also audit the randomized response mechanism \cite{warner1965randomized} with our method, and compare with Steinke et al.'s \cite{steinke2023privacy}. Since this experiment is not implemented by Mahloujifar et al. \cite{mahloujifar2024auditing}, we only compare with Steinke et al.'s \cite{steinke2023privacy}. This is not an issue, as we will see that our method already gives tight results. The setup follows Xiang et al.'s \cite{xiang2025privacy} implementation: 

For this experiment, $b$ is an $n$-bit vector to be guessed. The randomized response mechanism maps each bit $b_i$ as follows: if $b_i=0$, it outputs $0$ w.p. $\frac{(1-\delta)\mathrm{e}^\varepsilon}{1+\mathrm{e}^\varepsilon}$, $1$ w.p. $\frac{(1-\delta)}{1+\mathrm{e}^\varepsilon}$, and $2$ w.p. $\delta$; if $b_i=1$, it outputs $1$ w.p. $\frac{(1-\delta)\mathrm{e}^\varepsilon}{1+\mathrm{e}^\varepsilon}$, $0$ w.p. $\frac{(1-\delta)}{1+\mathrm{e}^\varepsilon}$, and $3$ w.p. $\delta$. This mechanism satisfies $(\varepsilon,\delta)$-DP. The guess $\hat{b}_i$ is made based on the observed output: \textit{we guess $0$ if the output is $0$ or $2$ and we guess $1$ if the output is $1$ or $3$}

The goal of this experiment is to show that for some large $\delta$ value, Steinke et al.'s \cite{steinke2023privacy} can fail to even give any meaningful result, while our method can still give tight results.

We give results in Figure \ref{fig:gdp_audit_rr}. We see that our method still outperforms Steinke et al.'s \cite{steinke2023privacy} method in all cases. The most notable phenomenon is that when $\delta=10^{-2}$, Steinke et al.'s \cite{steinke2023privacy} method can not give any meaningful result for $n>10^{2}$, while our method can still give a tight audit result. This is because the bound given by Steinke et al.'s \cite{steinke2023privacy} method includes a bad factor of $\mathcal{O}(n\delta)$ in the tail probability: once $n$ or $\delta$ is large, the tail bound becomes trivial, and Steinke et al.'s method then fails to give any meaningful result. 

\vspace{\smgbs}
\subsection{Audit Sub-sampled Gaussian Mechanism}\label{sec:audit_SMG}
\vspace{\smgbs}
We compare with Steinke et al.'s \cite{steinke2023privacy} method on the sub-sampled Gaussian mechanism (SGM). The SGM is a variant of the Gaussian mechanism where there is an additional layer of randomness. The mechanism is described as follows:
\begin{equation}\label{equ:sgm}
    \mathcal{M}(b)=\mathbf{Sub}_q(b)+\mathcal{N}(0,\sigma^2\mathbb{I}^n)
\end{equation} 
where the sub-sampling function $\mathbf{Sub}_q$ operates independently on each bit inside $b$ and keeps the original value w.p. $q$ and sets it to $0$ w.p. $1-q$. The guessing strategy is the same as that of the Gaussian mechanism. We use ``$1/\sigma$-GDP, $q$'' to denote the above mechanism.

This additional layer of randomness makes privacy auditing relatively hard even in the case of audit by multiple runs \cite{nasr2023tight}. However, we are able to reach tight audit results even in one run.
Before presenting our results, we first give the base distribution pair in Figure \ref{fig:dif_PQ_sgm}. The trade-off function's derivation follows Dong et al.'s \cite{dong2019gaussian} general construction, and we extract the $P,Q$ base distribution pair for our purpose.

\textit{The guessing strategy is the same as that audit Gaussian mechanism in the special case.} Results are presented in Figure \ref{fig:gdp_audit_sgm}. We compare our method with Steinke et al.'s \cite{steinke2023privacy} method. We can reach tight results in one run, even when there is sub-sampling. As mentioned before, the $\delta$ parameter can have a negative impact on Steinke et al.'s \cite{steinke2023privacy} method, so we set $\delta=10^{-7}$ to ensure that the impact is not significant. In this experiment, \cite{steinke2023privacy}'s method can only give much weaker results than ours. More results regarding different $q$ are provided in Appendix \ref{app:dpsgd}, which also gives similar conclusions.

% \vspace{\smgbs}
\subsection{Parameter Ablations}\label{sec:ablation}

\notbf{P-value.} The right-hand side of Equation \eqref{equ:best_tail_bound} serves as our p-value for the privacy audit. We will compare the p-value given by our method with that given by Steinke et al.'s \cite{steinke2023privacy} method. The targeted algorithm satisfies $(\varepsilon,\delta)$-DP, and we manually vary the guessing accuracy to compare the performance.

Results are presented in Figure \ref{fig:p_value}. We can see that when $\delta=10^{-2}$, our method gives a p-value that is smaller, while we can still reject the null hypothesis. For example, when the number of wrong guesses is $10$, our method will reject the null hypothesis while Steinke et al.'s \cite{steinke2023privacy} method will not.
For some larger value of $\delta$, e.g., $\delta=10^{-1}$, our method still has robust performance, but Steinke et al.'s \cite{steinke2023privacy} method can not reject the null hypothesis at any number of wrong guesses. In such a case, they will produce a trivial privacy lower bound, i.e., $\varepsilon_l=0$, just like results shown in Figure \ref{fig:gdp_audit_rr} when $\delta=10^{-2}$.

% \vspace{\MargBtParag}
\notbf{Different $n$ at fixed $r$ released guesses.} In this experiment, we simulate the cases where we only release guesses with a fixed number $r$ but vary the total number of guesses $n$. This is to investigate the sensitivity of the audit performance under different $n$. Results are presented in Figure \ref{fig:fix_r_dif_n}. 

Suppose $c$ is the number of correct guesses out of $r=1000$ released guesses. We can see that with $n$ increasing, the guessing accuracy $c/r$ in auditing the Gaussian mechanism also increases. This is because more extreme values are likely to be observed when $n$ is larger, and extreme values are more likely to be guessed correctly for the Gaussian mechanism. We can also see that our method is robust to the change of $n$ and can still give results close to the upper bound at fixed $r=1000$. In contrast, Steinke et al.'s \cite{steinke2023privacy} method cannot give any meaningful result when $n$ is some large number.

% \vspace{\MargBtParag}
\notbf{Do we really have to trade off over $r$?} One important question is to ask why we should set $r<n$, i.e., only release a subset of guesses. The intuition given by previous work, Steinke et al. \cite{steinke2023privacy} is that by focusing on the most ``confident'' guesses, a better privacy audit result can be achieved, as it tends to increase the guessing accuracy. 

But it is notable that \cite{steinke2023privacy}'s method must trade off between the number of released guesses $r$ and the confidence level, i.e., the larger $r$ is, the guessing accuracy will decrease, although the confidence level will increase. This phenomenon is captured in Figure \ref{fig:gdp_audit_r_trade_off}, where the $\varepsilon$ lower bound given by Steinke et al. first increases and then decreases as $r$ increases. And we can also see that the guessing accuracy indeed tends to decrease as $r$ increases. We emphasize that a similar phenomenon is also observed in previous work \cite{mahloujifar2024auditing} for their method.

In contrast, using our method, Figure \ref{fig:gdp_audit_r_trade_off} shows that it may not be necessary to trade off between the number of released guesses $r$ and the confidence level: our $\varepsilon$ lower bound is monotonically increasing as $r$ increases. Such results give evidence that the confidence level is a more crucial factor in achieving better privacy audit results, and our results suggest that we can achieve better privacy audit results by just releasing more guesses.

We must point out that such a conclusion should be used with caution: if some guesses are indeed just random, we clearly should not release them. In practice, some guesses could be really bad (close to random guessing), such that they should not be released. However, analyzing how to set $r$ requires more assumptions that depend on the specific applications. Nevertheless, in our experiment, our method gives better lower bounds for any $r$.

% \vspace{-0.3cm}
\section{Conclusion}\label{sec:conclusion}
% \vspace{-0.3cm}
We provide an affirmative answer that tight privacy audit for general DP protocols can be achieved in a single run. With our refined framework, by modeling the target algorithm using the more expressive $f$-DP formulation, we conduct a theoretical analysis to characterize the fundamental implications of our framework. The behavior of releasing a subset of guesses is captured through the lens of order statistics, which we leverage to design a method capable of producing a privacy lower bound in just one run.

We integrate these components into an auditing method and demonstrate, through experiments, that our method achieves tight privacy lower bounds and outperforms prior work across multiple scenarios. Contrary to earlier findings, we find that it is possible to simplify parameter setup and avoid the need for a trade-off over the number of released guesses. Finally, our method can be deployed as an add-on to improve the privacy lower bounds in practical auditing workflows.
% \notbf{}
% \section*{Acknowledgements}

%% file: sp26_appendix.tex
\appendices

\section{Audit Implementation}\label{sec:audit_method}
\vspace{-0.15cm}
\notbf{Implementation.}  First, we set our null hypothesis that the targeted privacy algorithm $\mathcal{M}$ is some $f$-DP. We then run the audit experiment as described, and observe $u$ errors among the $r$ released bits. According to Theorem \ref{thm:best_tail_bound}, we can compute the probability that an algorithm with $f$-DP privacy guarantee would result in at most $u$ errors among the $r$ released guesses. If this probability is not greater than a chosen significance level (e.g., $0.05$), we then reject the null hypothesis and conclude with confidence (e.g., $1-0.05$) that $\mathcal{M}$ does not satisfy $f$-DP, which then gives a privacy lower bound. We revise the null hypothesis to reach the strongest privacy lower bound that can still be rejected. The lower bound is often described via the $(\varepsilon,\delta)$-DP form by converting from $f$-DP to $(\varepsilon,\delta)$-DP following Algorithm \ref{alg:fdp_to_eps_delta}.

\begin{algorithm}[!ht]
\caption{
$f$-DP to $(\varepsilon,\delta)$-DP \cite{dong2019gaussian,zhu2022optimal}
}\label{alg:fdp_to_eps_delta}

\begin{algorithmic}[1]
% \equsize
\renewcommand{\algorithmicrequire}{\textbf{Input:}}
\renewcommand{\algorithmicensure}{\textbf{Output:}}

\Require {$f$, trade-off function; $\delta$, privacy parameter}

\State Return $\infty$ if $\delta < 1-f(0)$
\State $\varepsilon=\inf\{a:f(x)\geq 1-\delta-\mathrm{e}^a x, \forall x\in[0,1]\}$

\Ensure $\max\{0, \varepsilon\}$

\end{algorithmic}
\end{algorithm}

\begin{table}[ht]
\centering
\begin{tabular}{ccccc}
\toprule
$n$ & $r$ & Target $\varepsilon$ & Steinke et al. \cite{steinke2023privacy} & \textbf{Ours} \\
\midrule
500  & 124 & 1 & $0.160\pm0.103$ & $\mathbf{0.339}\pm0.227$ \\
500  & 124 & 2 & $0.228\pm0.161$ & $\mathbf{0.499}\pm0.367$ \\
500  & 124 & 4 & $0.722\pm0.270$ & $\mathbf{1.704}\pm0.680$ \\
500  & 124 & 8 & $1.331\pm0.249$ & $\mathbf{3.250}\pm0.637$ \\
1000 & 250 & 1 & $0.165\pm0.093$ & $\mathbf{0.350}\pm0.207$ \\
1000 & 250 & 2 & $0.429\pm0.135$ & $\mathbf{1.973}\pm0.314$ \\
1000 & 250 & 4 & $0.906\pm0.089$ & $\mathbf{2.469}\pm0.228$ \\
1000 & 250 & 8 & $1.460\pm0.188$ & $\mathbf{3.589}\pm0.480$ \\
\bottomrule
\end{tabular}
\caption{
\RevisionChangeStart
More results on the DP-SGD experiments for $r=n/4$
\RevisionChangeEnd
}
\vspace{-0.2cm}
\label{tab:more_dpsgd4rn}
\end{table}

\begin{table}[ht]
\centering

\begin{tabular}{ccccc}
\toprule
$n$ & $r$ & Target $\varepsilon$ & Steinke et al. \cite{steinke2023privacy} & \textbf{Ours} \\
\midrule
500  & 250 & 1 & $0.103\pm0.078$ & $\mathbf{0.281}\pm0.224$ \\
500  & 250 & 2 & $0.225\pm0.113$ & $\mathbf{0.645}\pm0.335$ \\
500  & 250 & 4 & $0.611\pm0.156$ & $\mathbf{1.888}\pm0.520$ \\
500  & 250 & 8 & $1.039\pm0.137$ & $\mathbf{3.335}\pm0.466$ \\
1000 & 500 & 1 & $0.163\pm0.105$ & $\mathbf{0.458}\pm0.324$ \\
1000 & 500 & 2 & $0.336\pm0.074$ & $\mathbf{2.185}\pm0.232$ \\
1000 & 500 & 4 & $0.688\pm0.046$ & $\mathbf{2.342}\pm0.155$ \\
1000 & 500 & 8 & $1.104\pm0.109$ & $\mathbf{4.557}\pm0.373$ \\
\bottomrule
\end{tabular}

\caption{
\RevisionChangeStart
More results on the DP-SGD experiments for $r=n/2$
\RevisionChangeEnd
}
\vspace{-0.2cm}
\label{tab:more_dpsgd2rn}
\end{table}

\begin{table}[ht]
\centering
\begin{tabular}{c c c c c}
% \hline
\toprule
 & $n=10^2$ & $n=10^4$ & $n=10^6$ & $\varepsilon_T$ \\
\toprule
$q=0.1$ & $<0.01, 0.03>$ & $<0.09, 0.29>$ & $<0.12, 0.32>$ & 0.39 \\
$q=0.9$ & $<0.02, 0.36>$ & $<0.52, 1.74>$ & $<0.52, 1.83>$ & 1.88 \\
\toprule
\end{tabular}

\caption{
\RevisionChangeStart
More results on the sampling rate $q$ in the auditing sub-sampled Gaussian mechanism. In each cell, results are given by $<$\cite{steinke2023privacy}'s, ours$>$. $\varepsilon_T$ means the true privacy parameter.
\RevisionChangeEnd
}
\vspace{-0.2cm}
\label{tab:more_q}
\end{table}

\begin{figure}[!t] 
    \centering
    % \subfloat[]
    {
    \includegraphics[width=1\linewidth]{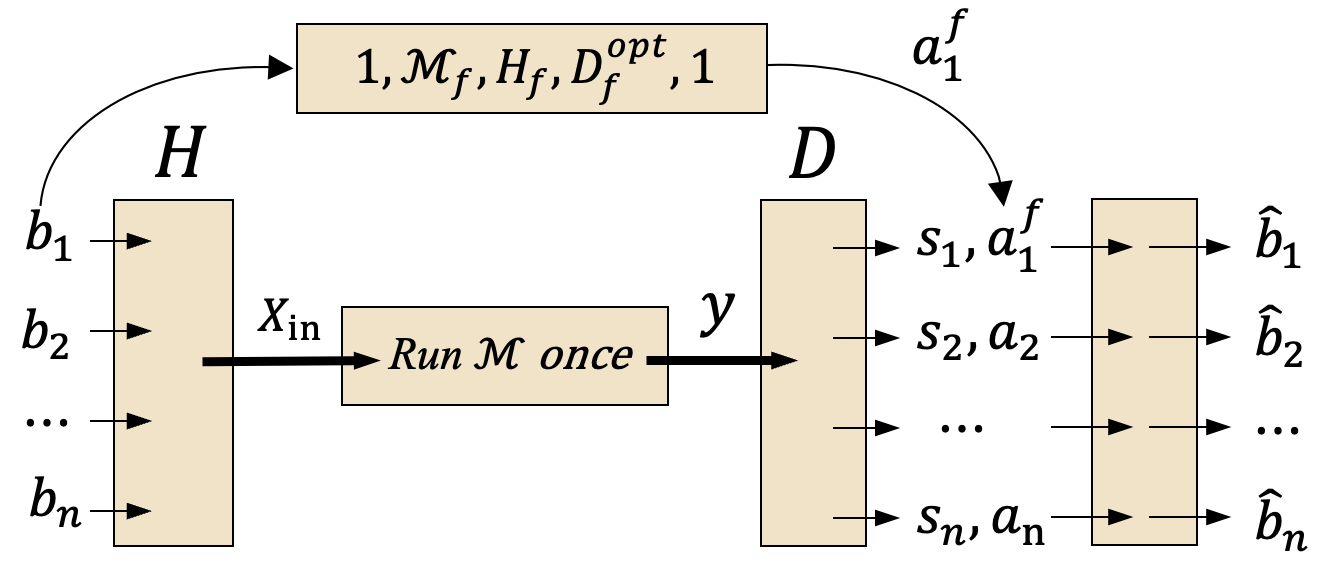}
    }
    \vspace{\fmgftoc}
    \caption{
    Pictorial description of the framework $(n, \mathcal{M}_1, \mathbf{H}_1, \mathbf{D}_1, r)$ where the guessed bit $a_1$ is overwritten by the guessed bit $a_1^f$ of the $f$-DP channel $(1, \mathcal{M}_f, \mathbf{H}_f, \mathbf{D}^{opt}_f, 1)$. Note that for the independent $f$-DP channel, $\mathcal{M}_f$ has the same privacy guarantee as $\mathcal{M}$, i.e., $f$-DP. 
    }
    \label{fig:first_better_channel} 
    \vspace{\fmgctom}
    \end{figure}

\section{More Experiments and Details}\label{app:dpsgd}

% \vspace{-0.05cm}
\noindent\textbf{Experimental Details of DP-SGD}. We use CIFAR10 dataset and “white-box” setting. During standard DP-SGD training with default setup of \textit{Adam} optimizer, all intermediate states are released, and we insert one-hot gradient canaries (“Dirac canaries”). The guessing strategy is also identical to \cite{steinke2023privacy}: we set $n=10000, r= 5000$, we find 2500 coordinates having the smallest value to be guessed with 0, and another 2500 having the largest value to be guessed with 1.
The only difference is the network: we use Resnet14 (from \cite{xiang2025privacy}’s GitHub) instead of Wide-Resnet used in \cite{steinke2023privacy} due to GPU-out-of-memory. But we argue this discrepancy is irrelevant to our conclusion because we directly control the gradient in the “white-box” setting. It hardly depends on any network. Most importantly, to further rule out any factors leading to unfair comparison, our method and \cite{steinke2023privacy}'s are applied to the same set of released guesses. For more result on diffent $r$ and $n$, results are given in Table \ref{tab:more_dpsgd4rn} and \ref{tab:more_dpsgd2rn}

% \vspace{-0.05cm}
% \vspace{0.1cm}
\noindent\textbf{More results regarding different sampling rate $q$ in auditing sub-sampled Gaussian mechanism.} Results are shown in Table \ref{tab:more_q}. We see auditing becomes more challenging as $q$ becomes small. We think it is because guessing becomes harder accordingly. Nevertheless, we can still obtain near tight result when n is large and our result are always much better than \cite{steinke2023privacy}.

\section{Proofs}
\subsection{Proof of Lemma \ref{lem:optimal_decoder_f_dp_channel}}\label{sec:proof_optimal_decoder_f_dp_channel}

\begin{proof}
    Bit error in two situations: the false positive case where $B=0$ but $\hat{B}=1$, and the false negative case where $B=1$ but $\hat{B}=0$. Let the false positive rate be $\alpha$ and the false negative rate at best is $f(\alpha)$. Given $B\sim\mathbf{Bernoulli}(1/2)$. The bit error probability is
    $$
    e(\alpha)=\frac{\alpha+f(\alpha)}{2}
    $$
    since $f$ is convex, then $e(\alpha)$ is convex, a point $\alpha^{opt}$ is a global minimum of $e(\alpha)$ if and only if 0 is contained in the subgradient of $e(\alpha)$ at $\alpha^{opt}$. Equivalently, at $\alpha^{opt}$, $-1$ is a subgradient of $f$.

\end{proof}
\sheperdrevisionStart
\subsection{Proof of Lemma \ref{lem:error_prob}}\label{sec:proof_error_prob}

\begin{proof}
When $P(y)\geq Q(y)$, the optimal decoder outputs $0$, error occurs when $B=1$, hence 
$$
\Pr(E=1\mid y)=\Pr(B=1\mid y)=\frac{1}{1+P(y)/Q(y)}
$$
similarly, when $Q(y)> P(y)$
$$
\Pr(E=1\mid y)=\Pr(B=0\mid y)=\frac{1}{1+Q(y)/P(y)}
$$
So 
$$
\Pr(E=1\mid y)=\frac{1}{1+\max(P(y)/Q(y),Q(y)/P(y))}
$$ which is just an equivalent form for Equation \eqref{equ:posterior_best_bit_error}

% \frac{1}{1+ \exp(\left|\mathcal{L}_{P/Q}(y)\right|)
    
\end{proof}

\sheperdrevisionEnd

\subsection{Proof of theorem \ref{thm:trans_dominated_by_ind_f_dp_channel}}\label{sec:proof_trans_dominated_by_ind_f_dp_channel}

\RevisionChangeStart
\begin{assumption}[Also enforced as the audit design principle]\label{ass:no_cross_adv}
    First, order the secret bits arbitrarily; for every $k\in[n]$ and every other realized error bit
    $e_{-k}=(e_1,\dots,e_{k-1}, e_{k+1},\dots, e_n)$,
    \begin{equation}\label{equ:seq_cond}
        \pr{}{E_k=1 \mid E_{-k}=e_{-k}}\;\geq\;e^{opt}.
    \end{equation}
    where $e^{opt}=\frac{\alpha^{opt} +f(\alpha^{opt})}{2}$ is the best error probability achievable shown in Equation \eqref{equ:lowest_bit_error_prob}.
\end{assumption}
\notbf{Remark.} The design principle can be enforced, and it also serves as the assumption for this Theorem. 

1) This principle does not require all $E_i$ to be independent from each other (which is a stronger assumption and corresponds to the ``\textbf{Special Case}'' in section \ref{sec:comparison_special} )

2) This principle just states a natural fact: we should not ``cheat'' in privacy audit; knowing how other guessing is correct or not should never make the current bit easier to guess than the best achievable an $f$-DP channel allows. In the following, we give a counter-example showing a contrived and strange audit design violates such a principle and will make our theorem fail.

\notbf{Counter-example.} Suppose we set $\mathcal{M}$ to be the mechanism that releases all secret bits w.p. $\delta$ and outputs $\emptyset$ w.p. $1-\delta$. The best guessing strategy is to guess the same bits we get if we see something non-empty and take random guesses if we see $\emptyset$. In this case, knowing many bits other than $k-th$ that are guessed correctly will almost surely give us the information that the audited algorithm's output is non-empty; thus, we know $k-th$ bits must be guessed correctly. 

\notbf{Discussion.} Theorem \ref{thm:trans_dominated_by_ind_f_dp_channel} no longer applies to the above case because the probability of making $0$ error is $\delta$ plus some small binomial probabilities (when the output is $\emptyset$ and all random guessing happens to be correct), but our Theorem 1 only gives that making $0$ error has some small binomial probabilities. However, in real-world privacy audits, such a contrived outlier almost does not exist. 

All of the experiments we have done comply with the audit principle: all of the DP algorithms are not like the contrived one, and any inserted canary does not expose other canaries. We must also note that all of the audit experiments done in previous work like \cite{xiang2025privacy,steinke2023privacy,nasr2021adversary,nasr2023tight} also comply with the principle.

\RevisionChangeEnd

\begin{proof}
    First, based on $( n, \mathcal{M}, \mathbf{H}; \mathbf{D}, r)$, we will construct a framework $(n, \mathcal{M}_1, \mathbf{H}_1, \mathbf{D}_1, r)$ and show that 
    $$
    (n, \mathcal{M}, \mathbf{H}, \mathbf{D}, r) \preceq (n, \mathcal{M}_1, \mathbf{H}_1, \mathbf{D}_1, r).
    $$
    note that the new framework only differs in $\mathcal{M}_1$, $\mathbf{H}_1$ and $\mathbf{D}_1$. The high-level description for them is as follows: for input bit $b_1$, we construct a parallel "path" where $b_i$ is transmitting through a $f$-DP channel $(1, \mathcal{M}_f, \mathbf{H}_f, \mathbf{D}^{opt}_f, 1)$; then we let the guessed bit $a_1^f$ of the such $f$-DP channel to take over the position of the guessed bit $a_1$ of the original framework. We show $(1, \mathcal{M}_f, \mathbf{H}_f, \mathbf{D}^{opt}_f, 1)$ pictorially in Figure \ref{fig:first_better_channel}.

    Note that the ``take-over'' action does not affect the action of the original filtering action, because the filtering is based on the scores, and the original score $s_1$ is preserved. I.e., $a_1^f$ is released if and only if the original $a_1$ is released. Consequently, $a_1^f$ is a better guess (in terms of bit error) than the original $a_1$ because of the optimal decoder function $\mathbf{D}^{opt}_f$ in Lemma \ref{lem:optimal_decoder_f_dp_channel}. Quantitatively, define $\mathbf{event}_u$ the same as Equation \eqref{equ:event_u}, for the case where $a_1$ is not released, clearly we have
    \begin{equation}\label{equ:first_prob_dominance_1}
    \begin{aligned}
        \pr{B, \mathcal{M}}{\mathbf{event}_u}  = \pr{B, \mathcal{M}_1}{\mathbf{event}_u}
    \end{aligned}       
    \end{equation}
    as the rest of the output $b_i, \forall i>1$ remains unaffected.
    for the case where $a_1$ is released, we have
    \equienvs
    \begin{equation}\label{equ:first_prob_dominance_2}
    % \equsize
    \begin{aligned}
        &\pr{B, \mathcal{M}}{\mathbf{event}_u}\\
        =& \pr{B, \mathcal{M}}{E_1\leq u - \mathbf{SUM}(\{E_i:i>1,E_i\neq\bot\})}\\
        =& \underset{E_{>1}}{\mathbb{E}}\left[\pr{B_1, \mathcal{M}}{E_1\leq u - \mathbf{SUM}(\{E_i:i>1,E_i\neq\bot\})|E_{>1}}\right]\\
        \overset{\mathbf{1}}{\leq}&\underset{E_{>1}}{\mathbb{E}}\left[\pr{B_1, \mathcal{M}_1}{E_1\leq u - \mathbf{SUM}(\{E_i:i>1,E_i\neq\bot\})|E_{>1}}\right]\\
        =&\pr{B, \mathcal{M}_1}{E_1\leq u - \mathbf{SUM}(\{E_i:i>1,E_i\neq\bot\})}\\
        =&\pr{B, \mathcal{M}_1}{\mathbf{event}_u}
    \end{aligned} 
    \end{equation}
    \equienve
    Inequality $\mathbf{1}$ is because 1) 
    \RevisionChangeStart
    Assumption \ref{ass:no_cross_adv} (or the enforced principle), ensures that knowing how other guessing is correct or not can’t make the current bit easier than the best achievable an $f$ -DP channel allows (which is the additional independent channel $\mathcal{M}_1$ has).
    \RevisionChangeEnd
    2)  the randomness inside $(1, \mathcal{M}_f, \mathbf{H}_f, \mathbf{D}^{opt}_f, 1)$ is independent from everything else, and 3) for $0<v\leq w<1$ and $V\sim\mathbf{Bernoulli}(v), W\sim\mathbf{Bernoulli}(w)$, we have 
    $$
    \pr{}{W\leq u} \leq \pr{}{V\leq u}, \forall u
    $$

    Then, we can do this inductively: for $(n, \mathcal{M}_k, \mathbf{H}_k, \mathbf{D}_k, r)$, there are $k$ independent parallel $f$-DP channels $(1, \mathcal{M}_f, \mathbf{H}_f, \mathbf{D}^{opt}_f, 1)$ in addition to the original framework $(n, \mathcal{M}, \mathbf{H}, \mathbf{D}, r)$, and the guessed bit $a_i, \forall i =1,..,k$ is \textbf{overwritten} by the guessed bit $a_i^f$ of the $f$-DP channels. By the same reasoning shown in the above and the fact that randomness in each $f$-DP channel is independent from each other and independent from everything else, we have
    $$
    \pr{B, \mathcal{M}_k}{\mathbf{event}_u} \leq \pr{B, \mathcal{M}_{k+1}}{\mathbf{event}_u}
    $$
    Finally, we have 
    $$
    \pr{B, \mathcal{M}}{\mathbf{event}_u} \leq \pr{B, \mathcal{M}_n}{\mathbf{event}_u}
    $$
    i.e,
    $$
    (n, \mathcal{M}, \mathbf{H}, \mathbf{D}, r) \preceq (n, \mathcal{M}_n, \mathbf{H}_n, \mathbf{D}_n, r).
    $$
    There is one step remaining, i.e., we need to show the following holds:
    $$
    (n, \mathcal{M}_n, \mathbf{H}_n, \mathbf{D}_n, r) \preceq (n, \mathcal{M}', \mathbf{H}', \mathbf{D}', r).
    $$
    It is not hard to see that for all guessed bits from those $f$-DP channels, the key difference between $(n, \mathcal{M}_n, \mathbf{H}_n, \mathbf{D}_n, r)$ and $(n, \mathcal{M}', \mathbf{H}', \mathbf{D}', r)$ is during the filtering step: for $(n, \mathcal{M}_n, \mathbf{H}_n, \mathbf{D}_n, r))$, the filtering is based on scores computed from the original framework $(n, \mathcal{M}, \mathbf{H}, \mathbf{D}, r)$; while for $(n, \mathcal{M}', \mathbf{H}', \mathbf{D}', r)$, the filtering is based on the scores $\left|\mathcal{L}_{P/Q}(y_i)\right|$ from itself where $y_i$ is the output of $\mathcal{M}_f$ from the $i$-th paralell $f$-DP channel.

    Intuitively, in terms of making bit errors, filtering based on scores computed from the original framework is suboptimal (which can release bits with very low scores), and we show in the following that it is indeed the case.

    The remaining problem can be simplified to the following. Conditioned on $(y_1,y_2,\cdots,y_n)$ where each independent $y_i$ is the output of $\mathcal{M}_f$ from the $i$-th parallel $f$-DP channel. 
    \begin{equation}\label{equ:posterior_bit_prob_f_dp_channel_proof}
    % \large
        B_i|y_i = \left\{\begin{array}{ll}
            0 & w.p. \quad \frac{P(y_i)/Q(y_i)}{1+ P(y_i)/Q(y_i)}\\
            1 & w.p. \quad \frac{1}{1+ P(y_i)/Q(y_i)}\\
        \end{array}\right.
    \end{equation}
    Because of the optimal decoder $\mathbf{D}^{opt}_f$,
    \begin{equation}\label{equ:posterior_error_bit_prob_f_dp_channel_proof}
        \pr{}{E_i=1|y_i} = \frac{1}{1+ \exp(\left|\mathcal{L}_{P/Q}(y_i)\right|)}
    \end{equation}

    Then 
    $$
    E_i|y_i\sim \mathbf{Bernoulli}(\frac{1}{1+ \exp(\left|\mathcal{L}_{P/Q}(y_i)\right|)}).
    $$ 
    % The question then becomes: how to choose $r$ bits among $\{a_i^G:i\in[n]\}$ so that $\pr{}{\mathbf{SUM}(\{E_i: E_i\neq\bot\})\leq u}$ is maximized. 
    And $E_i|y_i$ are independent from each other. It is immediately to see that the optimal choice is to choose the $r$ bits with the largest score $\left|\mathcal{L}_{P/Q}(y_i)\right|$ (corresponding to lowest $r$ bit error probabilities), so that $\pr{}{\mathbf{SUM}(\{E_i|y_i: \hat{B}_i|y_i\neq\bot\})\leq u}$ is maximized. Average over all $y_i$ which are independent sample from each other, $\pr{}{\mathbf{SUM}(\{E_i: \hat{B}_i\neq\bot\})\leq u}$ is maximized.
    
    This is because of the simple fact of the dominance preservation of the sum of independent Bernoulli random variables:
    for $0\leq v_1,\cdots,v_r \leq 1$, $0\leq w_1,\cdots,w_r \leq 1$, $v_i\leq w_i, \forall i\in[r]$ and $V_i\sim\mathbf{Bernoulli}(v_i)$, $W_i\sim\mathbf{Bernoulli}(w_i)$, then
    $$
    \pr{}{\sum W_i\leq u} \leq \pr{}{\sum V_i\leq u}
    $$
    Such can be easily proved by first showing that $\pr{}{ W_i\leq u} \leq \pr{}{V_i\leq u}$ and the sum of independent random variables preserves the dominance. So it gives
    $$
    \pr{B, \mathcal{M}}{\mathbf{event}_u} \leq  \pr{B, \mathcal{M}_n}{\mathbf{event}_u} \leq \pr{B, \mathcal{M}'}{\mathbf{event}_u}
    $$
    And we conclude
    $$
    (n, \mathcal{M}, \mathbf{H}, \mathbf{D}, r) \preceq (n, \mathcal{M}', \mathbf{H}', \mathbf{D}', r).
    $$

\end{proof}

\subsection{Proof of Lemma \ref{lem:gbar}}\label{sec:proof_gamma_bar}

\RevisionChangeStart
\begin{proof}
By marginalizing over $w$,
\begin{equation}
\begin{aligned}
    &\pr{}{E=1,\;U>w}\\
    =&\int_w^1 \pr{}{E=1\mid U=w'}\,dw'=\int_w^1\gamma(w')\,dw'\\
\end{aligned}
\end{equation}
note that $\pr{}{U>w}=1-w$, by Bayesian's rule,  dividing gives $\pr{}{E=1\mid U>w}=\bar{\gamma}(w)$. Equivalently, conditioned on
$\{U>w\}$ the quantile is uniform on $(w,1)$ with density $1/(1-w)$, so $\bar{\gamma}(w)$ is
the average of $\gamma$ over the band above $w$, not its value at any single point.

\end{proof}

\sheperdrevisionStart
\subsection{Proof of Lemma \ref{lem:markov_chain}}\label{sec:proof_markov_chain}

\begin{proof}
Intuitively, conditioned on $Z_{(i)}=z_i$, we know $Z_{(i+1)},...,Z_{(n)}$ are greater than $z_i$, this is exactly the information we get if conditioned on $Z_{(1)}=z_1,...,Z_{(i)}=z_i$, hence the claim holds. In the following, we give the formal proof. Assume that the density $g$ is continuous, so the order statistics are distinct with probability one. Let
$$
Z_{(1)}<Z_{(2)}<\cdots<Z_{(n)}
$$
denote the order statistics of the i.i.d.\ random variables
$Z_1,\ldots,Z_n$.
$$
n!\prod_{j=1}^n g(z_j),
\qquad
z_1<z_2<\cdots<z_n.
$$
Fix $i\in{1,\ldots,n-1}$. We first derive the conditional distribution of the upper block
$$
(Z_{(i+1)},\ldots,Z_{(n)})
$$
given
$$
Z_{(1)}=z_1,\ldots,Z_{(i)}=z_i.
$$
By the definition of conditional density,
$$
\begin{aligned}
&g_{Z_{(i+1)},\ldots,Z_{(n)}
\mid Z_{(1)},\ldots,Z_{(i)}}
(z_{i+1},\ldots,z_n
\mid z_1,\ldots,z_i)
\\
&=
\frac{
g_{Z_{(1)},\ldots,Z_{(n)}}(z_1,\ldots,z_n)
}{
g_{Z_{(1)},\ldots,Z_{(i)}}(z_1,\ldots,z_i)
}
\end{aligned}
$$
Note that 

$$
\begin{aligned}
    &g_{Z_{(1)},\ldots,Z_{(i)}}(z_1,\ldots,z_i)\\
=&\frac{n!}{(n-i)!}
\left(1-G(z_i)\right)^{n-i}
\prod_{j=1}^i g(z_j),
\end{aligned}
$$
for
$$
z_1<\cdots<z_i.
$$
this is because, after fixing $z_1,\ldots,z_i$, the remaining $n-i$ observations must all be larger than $z_i$, and their total probability is
$$
\left(1-G(z_i)\right)^{n-i}.
$$
Therefore,
$$
\begin{aligned}
&g_{Z_{(i+1)},\ldots,Z_{(n)}
\mid Z_{(1)},\ldots,Z_{(i)}}
(z_{i+1},\ldots,z_n
\mid z_1,\ldots,z_i)
\\
=&
\frac{(n-i)!\prod_{j=i+1}^n g(z_j)}
{\left(1-G(z_i)\right)^{n-i}},
\end{aligned}
$$
for
$$
z_i<z_{i+1}<\cdots<z_n.
$$
Crucially, this expression depends on
$(z_1,\ldots,z_i)$ only through the single value $z_i$. Hence
$$
\begin{aligned}
&g_{Z_{(i+1)},\ldots,Z_{(n)}
\mid Z_{(1)},\ldots,Z_{(i)}}
(z_{i+1},\ldots,z_n
\mid z_1,\ldots,z_i)
\\
&=
g_{Z_{(i+1)},\ldots,Z_{(n)}
\mid Z_{(i)}}
(z_{i+1},\ldots,z_n
\mid z_i).
\end{aligned}
$$
Therefore,
$$
(Z_{(i+1)},\ldots,Z_{(n)})
\perp
(Z_{(1)},\ldots,Z_{(i-1)})
\mid Z_{(i)},
$$
which proves that
$$
(Z_{(1)},\ldots,Z_{(n)})
$$
forms a Markov chain. Let 
$$
g_z^+(z)=\frac{g(x)}{1-G(z)}
\mathbf{1}_{\{x>z\}}.
$$
which the density of a draw from $f$ conditioned on being larger than $z$ (the density is re-weighted by $1-G(z)$).
Now consider $n-i$ i.i.d.\ random variables with density $g_z^+$. Their order statistics have joint density
$$
(n-i)!\prod_{j=i+1}^n g_z^+(z_j),
\qquad
z<z_{i+1}<\cdots<z_n.
$$
Substituting the definition of $g_z^+$ gives
$$
\frac{(n-i)!\prod_{j=i+1}^n g(z_j)}
{\left(1-G(z)\right)^{n-i}},
\qquad
z<z_{i+1}<\cdots<z_n.
$$
This is exactly the conditional density derived above. Therefore,
conditioned on $Z_{(i)}=z$,
$$
(Z_{(i+1)},\ldots,Z_{(n)})
$$
has the same distribution as the order statistics of $n-i$ i.i.d.\ samples drawn from $g$ truncated to $(z,\infty)$.
% Thus, conditioned on $Z_{(i)}=z$, the upper block is independent of the lower block
% $$
% (Z_{(1)},\ldots,Z_{(i-1)})
% $$
% and depends on the past only through $Z_{(i)}$.
\end{proof}

\sheperdrevisionEnd
\subsection{Proof of Lemma \ref{lem:cut}}\label{sec:proof_cut}
\begin{proof}
Note that $U_i=F_Y(Y_i)$, which are i.i.d.\ uniform on $[0,1]$. This is well known, as any continuous random variable will be transformed to be uniform on $[0,1]$ by its c.d.f. The released guesses are exactly those with the $r$ largest scores,
i.e.\ the order statistics $U_{(n-r+1)},\dots,U_{(n)}$, and the cut is $W=U_{(n-r)}$.

For $w\in(0,1)$, $U_{(n-r)}\leq w$ holds if and only if at least
$n-r$ of the $U_i$ are at most $w$, so
$$
\pr{}{W\leq w}=\pr{}{\mathbf{Binomial}(n,w)\geq n-r}
$$ Intuitively, each $U_i$ is smaller than $w$ w.p. $w$, so we need more than $n-r$ of such. Differentiating in $w$ gives the
density 
$$
\frac{n!}{(n-r-1)!\,r!}w^{\,n-r-1}(1-w)^{r}
$$
which is that of
$\mathbf{Beta}(n-r,r+1)$.

Apply Lemma
\ref{lem:markov_chain} with $i=n-r$. Conditioned on $W=w$, the $r$ order statistics above
the cut are precisely the released ones and are distributed as the order
statistics of $r$ i.i.d.\ draws from the uniform law truncated to $(w,1)$. Note that we care about the sum $T$, i.e., order does not matter here anymore. Then we can treat each released one as i.i.d., conditioned $W=w$. By Lemma
\ref{lem:gbar}, each error indicator is Bernoulli with a parameter determined by its own
output; consequently, conditioned on the outputs, the indicators are independent. To compute such a parameter, we need to marginalize the $r$
i.i.d.\ truncated quantiles, each released indicator is therefore $\mathbf{Bernoulli}(\theta)$
with
$$
\theta=\E{V\sim\mathbf{Uniform}(w,1)}{\gamma(V)}=\frac{1}{1-w}\int_w^1\gamma(w')\,dw'=\bar{\gamma}(w),
$$
Their sum is $\mathbf{Binomial}(r,\bar{\gamma}(w))$,
which is Equation \eqref{equ:cut_law}.
\end{proof}

\subsection{Proof of Theorem \ref{thm:best_tail_bound}}\label{sec:proof_best_tail_bound}
\begin{proof}
By Lemma \ref{lem:cut} and a simple application of the law of total probability,
\begin{equation}\nonumber
\begin{aligned}
    &\pr{}{T\leq u}\\
=&\E{W}{\pr{}{T\leq u \mid W}}\\
=&\E{W}{\pr{}{\mathbf{Binomial}\big(r,\bar{\gamma}(W)\big)\leq u}}
\end{aligned}
\end{equation}

\end{proof}

\RevisionChangeEnd